\documentclass[lettersize,journal]{IEEEtran}
\usepackage{amsmath,amsfonts}
\usepackage{algorithmic}
\usepackage{algorithm}
\usepackage{array}
\usepackage[caption=false,font=normalsize,labelfont=sf,textfont=sf]{subfig}
\usepackage{textcomp}
\usepackage{stfloats}
\usepackage{url}
\usepackage{verbatim}
\usepackage{graphicx}
\usepackage[backend=biber,minnames=3,maxnames=3,isbn=false,url=false,eprint=false,giveninits=true,sorting=none,citestyle=numeric-comp]{biblatex}
\usepackage{booktabs}
\usepackage[switch,columnwise]{lineno}  
\usepackage{multirow}

\usepackage{caption}

\newcommand{\ith}[1]{#1^{(i)}}

\begin{document}

\title{A Fast and Generic Energy‑Shifting Transformer for Hybrid Monte Carlo Radiotherapy Dose Calculation}


\author{Chi-Hieu Pham, Didier Benoit, Vincent Bourbonne, Ulrike Schick, Dimitris Visvikis and Julien Bert

\thanks{This work did involve human subjects in its research. Because this study relied exclusively on retrospective clinical data, appropriate protection measures and anonymization procedures were implemented, and approval from the local ethics committee was obtained.}

\thanks{The authors are with LaTIM, INSERM-UMR1101, University of Brest, 29200 Brest, France (e-mail: julien.bert@univ-brest.fr). V. Bourbonne and U. Schick are also with the radiotherapy department of the Brest Hospital University, France.}
}

\markboth{IEEE Transactions}%
{Shell \MakeLowercase{\textit{et al.}}: A Sample Article Using IEEEtran.cls for IEEE Journals}

\maketitle

\begin{abstract}
We introduce a novel learning framework for accelerated Monte Carlo (MC) dose calculation termed Energy Shifting. This approach leverages deep learning to synthesize highly complex polyenergetic dose distributions directly from simple monoenergetic inputs under identical beam configurations. Unlike conventional denoising techniques, which rely on noisy low-count dose maps that compromise beam profile integrity, our method achieves superior cross-domain generalization on unseen datasets by integrating high-fidelity anatomical textures and source-specific beam similarity into the model's input space. Furthermore, we propose a novel 3D architecture termed TransUNetSE3D, featuring Transformer blocks for global context and Residual Squeeze-and-Excitation (SE) modules for adaptive channel-wise feature recalibration. Hierarchical representations of these blocks are fused into the network’s latent space alongside the primary dose-map parameters, allowing physics-aware reconstruction. This hybrid design outperforms existing U-Net and Transformer-based benchmarks in both spatial precision and structural preservation,  while maintaining the execution speed necessary for real-time use. Our proposed pipeline achieves a Gamma Passing Rate exceeding 98\% (3\%/3mm) compared to the MC reference, evaluated within the framework of a treatment planning system (TPS) using 6MV TrueBeam Linear Accelerator (LINAC) for prostate radiotherapy. These results offer a robust solution for fast volumetric dosimetry in adaptive radiotherapy.

\end{abstract}


\IEEEpeerreviewmaketitle

\section{Introduction}

\IEEEPARstart{D}{ose} calculation in external beam radiotherapy is predominantly based on deterministic models. These models are sufficiently accurate for routine clinical use and fast enough to support treatment plan optimization. However, they suffer from intrinsic limitations in their domain of validity. In specific configurations not originally anticipated—such as bone–muscle or air–tissue interfaces, errors can reach up to 13\% \cite{zaman_comparison_2019}, and even higher (25\%) in the presence of metallic implants \cite{paulu_evaluation_2017}. While such inaccuracies may remain acceptable for standard treatments, they become problematic for emerging strategies requiring higher precision, including current trends in dose escalation \cite{latifi_study_2014}. A second, more recent limitation arises with the development of adaptive radiotherapy, particularly online adaptive workflows \cite{qiu_online_2023}, where treatment plans must be recomputed within minutes. This paradigm shift demands dose calculation methods that are even faster than those currently deployed.

In contrast, Monte Carlo (MC) simulation is considered the gold standard in medical physics. Based on stochastic Markov chain transport, MC methods provide highly accurate and robust dose calculations without the validity constraints inherent to deterministic models. Their major drawback is computational cost, which remains prohibitive for routine clinical use. Nevertheless, MC simulation is widely used for validating deterministic algorithms and for research and development of treatment systems.

Given its accuracy and robustness, substantial efforts have been devoted to accelerating MC simulation with the hope of replacing deterministic models and enabling its use in online workflows. Despite decades of research, including variance reduction techniques \cite{baldacci_track_2015} and hardware acceleration using GPUs \cite{bert_geant4-based_2013, hissoiny_gpumcd:_2011}, MC methods, although significantly faster, remain insufficiently rapid for clinical deployment. Moreover, most variance reduction strategies and GPU based implementations are limited to low energy photon transport ($<$1 MeV) \cite{baldacci_track_2015, behlouli_improved_2018}, which restricts their applicability for external beam radiotherapy, where both high energy photons and secondary electrons must be accurately modeled.

In recent years, artificial intelligence, and deep learning (DL) in particular, has gained traction in dose calculation. Several studies have demonstrated its effectiveness in external radiotherapy \cite{kontaxis2020deepdose, zhang2023deep}, internal dosimetry \cite{lee2019deep}, and brachytherapy \cite{villa_fast_2022}. However, DL models trained to directly predict dose distributions behave similarly to deterministic algorithms already implemented in treatment planning systems. They can be viewed as “super deterministic” models that approximate Monte Carlo results more closely than classical algorithms, provided they are trained on large MC generated datasets \cite{villa2023fast}. Yet they inherit the same limitations in robustness and generalizability: a network trained for head and neck dosimetry will not achieve comparable accuracy for pelvic treatments, as DL models remain intrinsically tied to their training domain.
The objective of the present work is not to develop another end to end DL dose engine. Instead, we propose a hybrid method that combines fast Monte Carlo simulation with deep learning to leverage the strengths of both approaches: the accuracy and robustness of MC, and the computational speed of DL.
To this end, we introduce a new concept, Energy-Shifting. The idea is to perform a fast MC simulation restricted to low energy photons ($<$1 MeV), enabling the use of efficient variance reduction techniques and GPU acceleration. A DL model is then used to convert the resulting photon only dose distribution into a dose map corresponding to the full energy spectrum of a clinical radiotherapy beam, including contributions from secondary electrons, such as 6MV TrueBeam Linear Accelerator (LINAC) \cite{vazquezquino_monte_2012}. This approach aims to retain the physical rigor of MC while achieving computation times compatible with online adaptive radiotherapy. 

Inspired by the success of Vision Transformers (ViTs) \cite{dosovitskiy2020image} in medical image segmentation \cite{hatamizadeh2022unetr} and fast dose calculation \cite{pastor2023sub}, we propose TransUNET3D, a novel deep learning architecture that integrates Transformer blocks within a 3D U-shaped residual network. A key innovation of our approach is the enhancement of the residual units with Squeeze-and-Excitation (SE) blocks \cite{roy2018concurrent}, which enable adaptive channel-wise feature recalibration to prioritize the most informative dosimetric representations. In addition, unlike existing architectures that replace the convolutional encoder with Transformers \cite{hatamizadeh2021swin,hatamizadeh2022unetr}, we strategically preserve convolutional residual blocks to maintain the local connectivity and inductive bias necessary for high-resolution dose maps. These residual blocks are integrated in parallel with Transformer blocks, which capture the long-range global context of particle distribution within the dose maps. The feature representations from both streams are fused within the latent space of the network, effectively combining the strengths of both worlds: the localized precision of CNNs and the global relational awareness of Transformers. Furthermore, we adopt a patch-based training strategy, a technique proven effective in various medical imaging tasks \cite{hatamizadeh2022unetr,he2023swinunetr}, to mitigate the risk of overfitting and improve the representation of features across the learning model. This strategy allows the network to capture more effectively invariant physical and anatomical features of the image domain through local patches. To further ensure a physics-aware reconstruction, beam-specific parameters are embedded directly into the latent space, guiding the model toward superior performance in accelerated Monte Carlo dose calculation.

\section{Proposed method}

\subsection{Monte Carlo simulations}
To demonstrate the proposed method and benchmark it against state-of-the-art techniques, a series of Monte Carlo simulations were performed. Two CT image databases were employed, one containing brain localization data from a retrospective study conducted by the radiotherapy department at Brest University Hospital (CHU Brest), and another comprising pelvic data from the publicly available "prostate anatomical edges cases" dataset \cite{thompson_stress-testing_2023}. For the pelvic dataset, 127 CT images with prostate and planning target volume (PTV) masks were used (each image consisting of a voxel size of $2 \times 2 \times 2 \text{ mm}^3 $). For the head dataset, 22 CT images containing brain and PTV masks were included (voxel size of $2 \times 2 \times 2 \text{ mm}^3 $). The input matrix dimensions vary between the patient cohort.


External beam radiotherapy treatments were simulated using Monte Carlo simulations with Gate v10 \cite{sarrut_gate_2025}. For this a collimated beam was directed at patient phantoms derived from the CT scans. The geometry and treatment distances were configured according to the specifications of the TrueBeam Novalis system (VARIAN, Siemens, Germany). This simulation strategy enabled the creation of diverse training datasets. To introduce variability, each simulation run represented a distinct irradiation scenario. At the start of each run, a patient CT was randomly selected from the relevant database (head or pelvis). This was followed by random selection of a beam angle between 0$^\circ$ and 180$^\circ$, a beam collimation with a field opening between 10 mm and 80 mm at the PTV, and finally a random target position within the organ of interest (brain or prostate) using its mask. The objective was to generate a training dataset covering the full range of possible treatment beam configurations. For each simulation, all configuration parameters and the resulting dose map, recorded at the same resolution as the CT scans (2 mm$^3$), were stored.

For each anatomical site, simulations were conducted with two levels of statistical uncertainty, fast simulations using $2 \times 10^5$ particles with $40\%$ uncertainty at the PTV, and standard simulations using 20 million particles to achieve uncertainties below 5\% at the PTV (average 4.3\%). For each uncertainty setting, two beams were simulated. One accurately representing the 6 MV TrueBeam Novalis beam \cite{vazquezquino_monte_2012}, and another using a monoenergetic 500 keV photon beam. For the latter, the Track Length Estimator variance reduction method \cite{baldacci_track_2015} and GPU acceleration \cite{bert_geant4-based_2013} were applied. The choice of 500 keV ensured sufficient tissue penetration and provided attenuation information useful for AI-based generation of TrueBeam 6 MV dose maps, while remaining below 1 MeV to avoid electron transport simulation (mean free path shorter than voxel size). This constraint enabled the effective use of VRT and GPU acceleration.
In total, 1000 samples with different configurations were simulated for each of the training datasets: head and pelvis, each with 6 MV and 500 keV beams, and each under low- and high-uncertainty conditions.

\subsection{Beam transformation}

\begin{figure*}[ht!]
    \centering
    \includegraphics[width=1\linewidth]{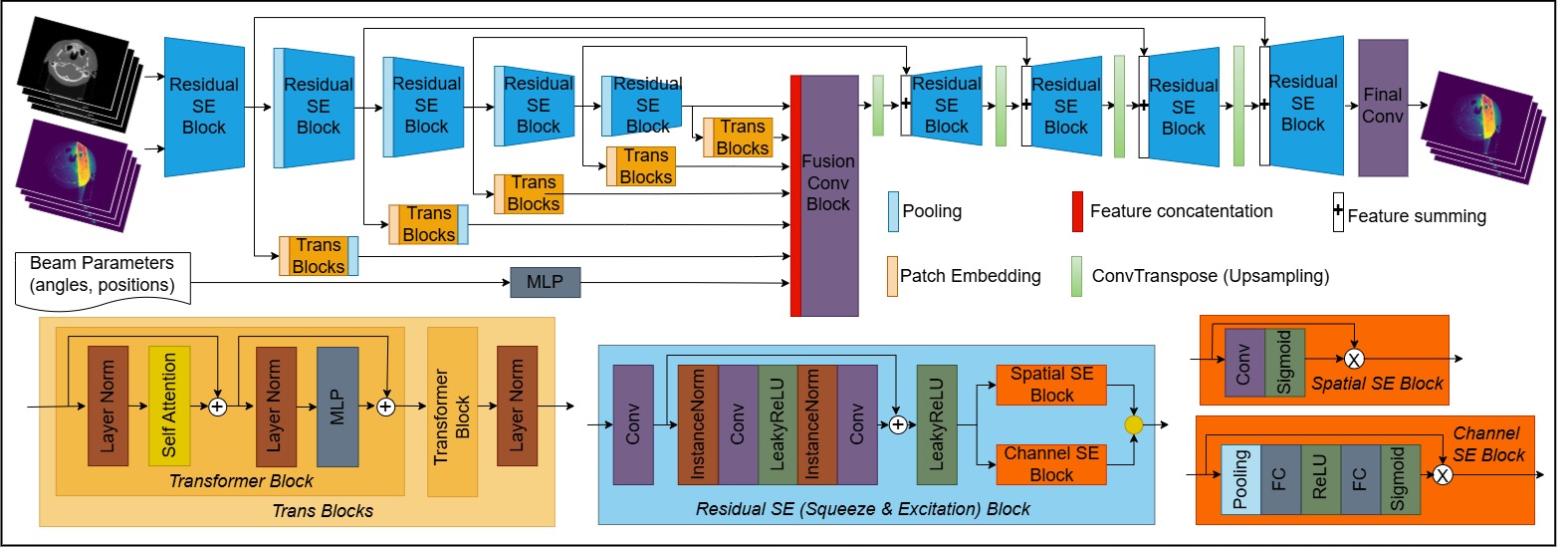}
    \caption{Overview of the architecture of the proposed TransUNetSE3D. }
    \label{fig:Overview}
\end{figure*}

We begin by addressing the problem of calculating the dose of MC-based radiotherapy in the previous work.  Let $V_{CT}$ represent a volumetric CT image and $Y$ denote a corresponding MC-based LINAC dose. $V_{CT}$ and $Y$ are of the same dimension (\textit{i.e.} $V_{CT} \in \mathbb{R}^{m \times n \times k}$ and $Y \in \mathbb{R}^{m \times n \times k}$). The objective of a learning-based method is to find a mapping $f_\theta$ from $V_{CT}$ to $\hat{Y}$ full doses, as defined by the following equation:

\begin{equation}
    \hat{Y} = f_\theta \left(V_{CT} \right) 
\end{equation}

where $\theta$ denotes the parameters of the mapping such as the weights and the biases of the convolutional neural networks. Recently, the deep UNet network \cite{ronneberger2015u, cciccek20163d} has become a crucial baseline architecture for medical image segmentation and fast dose calculation \cite{maier2018real,nguyen20193d, kontaxis2020deepdose,bai2021deep,maier2022real}. UNet is structured as two primary sub-networks: an encoder $f_{\text{Enc}}$ and a decoder $f_{\text{Dec}}$. The mapping of UNet can be described as $f_\theta (V_{CT}) = f_{\text{Dec}}( f_{\text{Enc}}(V_{CT}))$. The encoder maps the  CT image $V_{\text{CT}}$ into a highly compressed latent space representation $Z^{(J)}$ described as $Z^{(J)}=f_{\text{Enc}} \left( V_{CT} \right) $. This process involves a fixed number of sequential convolution blocks $f_{\text{Enc}}^{(j)}$ (where $j=1,\dots,J$), with each block typically including convolutional layers, non-linear activation (\textit{e.g.} Rectified Linear Unit - ReLU), and a downsampling operation (\textit{e.g.} max-pooling). The encoder is described as:
\begin{equation}
\begin{aligned}
    Z^{(1)} &= f_\text{Enc}^{(1)} \left( V_{CT} \right) \\
    Z^{(j)} &= f_\text{Enc}^{(j)} \left( Z^{(j-1)} \right) \ \text{for} \ j=2,\dots,J
\end{aligned}
\end{equation}

where $Z^{(j)}$ denotes the feature maps output by the $j^{\text{th}}$ block of the encoder. To enhance the non-linearity and feature representation capacity within each encoder block, we utilize a sequence of two convolution blocks. Each convolution block is composed of an order of operations: a normalization layer, followed by a convolutional layer, and concluding with an activation function (e.g., Rectified Linear Unit (ReLU) or Leaky ReLU). In recent work, the Instance Normalization layer \cite{ulyanov2016instance} has demonstrated superior performance, particularly in the context of image translation tasks. The output feature maps of each encoder block are typically downsampled using a max-pooling layer, effectively reducing their spatial dimensions by a factor of one-half relative to the preceding block.

The decoder $f_{\text{Dec}}$ then maps the final latent space features $Z^{(J)}$  back to the desired output $\hat{Y}=f_\text{Dec}(Z^{(J)})$ through a series of up-convolution blocks $f_{\text{Dec}}^{(t)}$ composed of an upsampling layer and two convolution blocks (where $t$ runs from $J-1$ down to $1$). Given $Z^{(J)}=D^{(J)}$, the decoding process is described by the following sequence :
\begin{equation}
\begin{aligned}
    D^{(t)} &= f_\text{Dec}^{(t)} \left( D^{(t+1)}, Z^{(t)}   \right) \ \text{for} \ t=J-1,\dots,1 \\
\end{aligned}
\label{equ_decoder}
\end{equation}
In each decoder step $t$, an up-convolution layer (or an upsampling layer) is first applied to the input features $D_{t+1}$ to double their spatial dimensions. The resulting features are then concatenated with the corresponding high-resolution residual feature maps $Z_t$ obtained via the skip connection from the encoder. The combined feature set is finally processed by convolution layers to produce the output $D_t$. In the final layer, the network uses a final convolution layer to reconstruct the estimated full dose $\hat{Y}$ as : $  \hat{Y} = f_{\text{FinalConv}}\left( D^{(1)} \right)$. However, a direct mapping from only the CT scans risks losing detail regarding the radiotherapy beam parameters. These parameters are essential for enabling learning methods to accurately preserve the beam characteristics and form in the output dose distribution. Consequently, the learning approach needs more comprehensive input data for effective reconstruction. To address this limitation and better leverage the network's representation learning capacity, previous work in \cite{villa2023fast} proposes to embed the parameters of the radiotherapy beam $\alpha$  directly into the latent space as $Z^{(J)} = f_\text{Enc}^{(J)} \left( Z^{(J-1)} , f_{\text{MLP}}(\alpha) \right)$, where $f_{\text{NN}}$ denotes a Multilayer Perceptron (MLP) that converts $\alpha$ to have the same number of channels and size as that of the latent space. The transformation of beam parameters facilitates the channel-wise concatenation of the beam-conditioning features with the latent representation. This strategy allows the network to explicitly learn the intricate characteristics of the radiotherapy beam, which may be obscured or difficult to infer from the CT scan alone. The problem of fast dose calculation can be reformulated as follows: $ \hat{Y} = f_\theta \left(V_{CT}, \alpha \right) $, where $\alpha \in \mathbb{R}^b$ denotes the $b$ acquisition parameters of simulated beam (such as angles and source position). 

However, as with many data-driven techniques, an encoder with only CT volumes as input that lacks sufficient feature representation cannot guarantee the preservation of the morphology and spatial texture of the target dose distribution. An alternative strategy for accelerated dose calculation involves predicting a high-sampling (HS) dose map $Y$ from a mutilmodal input such as PET/CT in \cite{lee2019deep}, or a rapidly simulated, low-sampling (LS) counterpart $Y_{LS}$ \cite{martinot2021high, neph2021deepmc, zhang2023deep, he2025deep}. Unlike HS references, LS maps are generated using a significantly lower particle count, resulting in inherent stochastic noise and diminished structural detail. This denoising-based approach such as \cite{peng2019mcdnet,zhang2023deep} treats the fast dose calculation problem as a restoration task, formulated as follows: $ \hat{Y} = f_\theta \left(Y_{LS} \right)$. Because LS dose maps typically exhibit poor structural definition of organs and tissue interfaces, they provide insufficient anatomical context for the network. In \cite{van2022novel, he2025deep}, to improve the encoder’s feature representation of absorbed dose across different tissue types, the uncertainty of dose map or CT scan can be concatenated with the LS dose maps to provide more spatial characteristics. However, the inherent uncertainty in raw dose maps often results in insufficient geometric information for accurate reconstruction, sometimes leading to redundant or ambiguous features. Therefore, this work focuses on concatenating CT volumes with the dose maps to guide the reconstruction process, expressed as: $ \hat{Y} = f_\theta \left(V_{CT}, Y_{LS} \right)$. 

Unlike denoising-based approaches, which are often limited by the loss of structural detail by noisy LS inputs, we propose a novel framework that utilizes high-sampling dose maps derived from a mono-energetic photon source instead of LS ones. Our objective is the rapid calculation of the full True-Beam dose $Y$ from mono-energetic photon dose maps $X$. This strategy ensures that the input and output share identical particle counts, allowing the model to mitigate the complexities of denoising and focus exclusively on the spectral translation between the two simulation strategies. Because mono-energetic beams are computationally inexpensive to simulate, we can maintain high execution speeds without sacrificing dosimetric fidelity. Furthermore, we incorporate beam delivery parameters (e.g., energy, incidence angle) as conditioning inputs to enhance the model’s generalization across diverse dataset. Thus, our proposed framework is expressed as: 
\begin{equation}
 \hat{Y} = f_\theta \left(V_{CT}, X, \alpha \right)    
\end{equation}

where concretely, $X$ denotes mono-energetic 500 keV photon beam and $\hat{Y}$ denotes the prediction of 6 MV TrueBeam beam $Y$. To optimize feature extraction and stabilize training against vanishing gradients, several works \cite{wolny2020accurate,lee2017superhuman, zhang2022plan},  employ residual learning within the U-Net blocks. The  $j^{th}$  encoder $f_{\text{ResidualEnc}}^{(j)}$ first computes a feature map from a convolution layer as $f_{\text{Conv}}^{(j)}(Z^{(j-1)})$ and then a residual sum of its this input and the output of a residual descriptor with two convolutional layers $f_{\text{ResidualDes}}^{(j)}$, followed by a ReLU, as described as :
\begin{equation}
\begin{aligned}
    Z^{(j)} &= f_{\text{ResidualEnc}}^{(j)}(Z^{(j-1)}) \\
    &=  f_{\text{ReLU}} \left( f^{(j)}_{\text{Conv}}(Z^{(j-1)}) + f_{\text{ResidualDes}}^{(j)}(f^{(j)}_{\text{Conv}}(Z^{(j-1)}) \right)
\end{aligned}
\end{equation}

In order to progressively reduce the spatial dimensions of the feature maps, a max-pooling layer is implemented before each encoder stage, with the exception of the first block. In the decoder, spatial resolution is recovered via transposed convolutions $f_{\text{ConvTranspose}}$, and the resulting feature maps are joined by summation with the skip-connection features $Z^{(t)}$ from the encoder. The fused feature maps then are represented by a decoder block as defined as : 

\begin{equation}
    D^{(t)} = f_{\text{ResdiualDec}}^{(t)} \left( f_{\text{ConvTranspose}}(D^{(t+1)}) + Z^{(t)} \right)
\end{equation}

where $f_{\text{ResdiualDec}}^{(t)}$ maintains a convolutional architecture  analogous to the encoder blocks $f_{\text{ResidualEnc}}^{(j)}$ to ensure consistent refinement of the characteristics (\textit{i.e.} the residual block denoted as $f_{\text{Resdiual}}$). While standard convolutions treat all channels within a local neighborhood equally, Channel Squeeze-and-Excitation (cSE) blocks facilitate feature recalibration by explicitly modeling inter-channel dependencies \cite{zhu2019anatomynet,roy2018concurrent}. This mechanism enhances the network's representation by identifying salient features (\textit{e.g.} distinguishing bone from soft tissue). A cSE block, $f_{\text{cSE}}(Z^{(i)})$, is defined as the element-wise multiplication of the input $Z^{(i)}$ and its channel-wise descriptor: $f_{\text{cSE}}(Z^{(i)}) = Z^{(i)} \times f_{\text{cDes}}(Z^{(i)})$, where $f_{\text{cDes}}$ encompasses a global average pooling layer followed by a bottleneck structure (fully connected layer, ReLU, fully connected layers and Sigmoid activations). Conversely, Spatial Squeeze-and-Excitation (sSE) identifies the locations of important features \cite{hu2018squeeze}. The sSE operation is expressed as $f_{\text{sSE}}(Z^{(i)}) = Z^{(i)} \times f_{\text{Sigmoid}}(f_{\text{Conv}}(Z^{(i)}))$. By integrating these into a concurrent Spatial and Channel SE (scSE) block \cite{roy2018concurrent}, the network utilizes a dual-attention mechanism:
\begin{equation}
    f_{\text{scSE}}(Z^{(i)}) = \max(f_{\text{sSE}}(Z^{(i)}), f_{\text{cSE}}(Z^{(i)}))
\end{equation}
Implementing the scSE block following a residual block creates a Residual SE block denoted $f_{\text{residualSE}}$, as described as : $ f_{\text{residualSE}} (Z^{(i)})= f_{\text{scSE}}(f_{\text{residual}}(Z^{(i)}) )$. This configuration refines feature maps and enhances gradient flow by emphasizing the most relevant information while suppressing noise \cite{wolny2020accurate}. 

Recently, Transformer architectures \cite{dosovitskiy2020image,liu2021swin} have demonstrated significant performance gains over standard convolutional blocks in feature representation, yielding superior results in various tasks. The methodology proposed in \cite{pastor2023sub} utilizes a 2D transformer architecture into the latent space of 3D UNet, where 3D dose feature maps are decomposed into 2D slices for model ingestion. While this approach benefits from the high-capacity global attention mechanisms inherent to Transformers, treating volumetric data as a sequence of independent 2D planes can lead to a loss of interslice spatial coherence. TransUNet \cite{chen2021transunet} proposed the integration of Transformer blocks \cite{dosovitskiy2020image} into the UNet bottleneck (\textit{i.e.} the encoder's output) to enhance the representation of latent space. This approach was extended to 3D architectures in \cite{chen23103d}, where Transformer blocks were also integrated into the decoder to improve the connectivity between the latent space and the decoding stages. Although advanced Transformer-based decoders could improve predictive accuracy, these models did not take advantage of Transformer-based feature representations across all encoder blocks. In contrast, Swin-Unet \cite{cao2022swin} replaces all convolutional blocks of the classic UNet with Swin Transformer blocks \cite{liu2021swin} to better capture patch dependencies within each encoder and decoder stage. However, this transition often overlooks the inherent advantages of residual blocks regarding feature representation and optimized gradient flow. Besides, SegFormer \cite{xie2021segformer} introduced a different hierarchical Transformer encoder to extract both coarse and fine features, coupled with a lightweight decoder for direct multi-level feature fusion. Although this reduces the parameter count, the method bypasses the capacity for multi-scale feature reuse typically found in traditional encoder-decoder skip connections such as UNet. On another hand, UNETR \cite{hatamizadeh2022unetr} proposes replacing the standard encoder of the UNet with a Vision Transformer (ViT) \cite{dosovitskiy2020image} while maintaining a convolutional decoder. Thus, the decoder extracts learned features from various stages of the ViT-based encoder, the well-established architecture, via skip connections. Building on this, SwinUNETR \cite{hatamizadeh2021swin,he2023swinunetr} further evolves the concept by employing a Swin Transformer \cite{liu2021swin} as the encoder. This hierarchical approach is better suited for handling multi-scale 3D medical data, as it effectively captures long-range patch dependencies at multiple resolutions. Nevertheless, this transition typically bypasses the benefits of residual learning found in established U-Net frameworks, which play a vital role in stabilizing gradient propagation and refining feature hierarchies. \\

In this work, we propose a novel 3D architecture termed TransUnetSE3D. This framework integrates Transformer blocks and Residual Squeeze-and-Excitation (SE) blocks within a 3D U-Net backbone to maximize global context awareness and local feature recalibration. By leveraging the self-attention mechanism within the Transformer blocks, the model can capture long-range dependencies across the feature maps instantaneously, characterizing relationships between voxels regardless of their spatial distance. Thus, we propose embedding a Transformer block to process the feature output of each Residual SE block within the encoder. Each stage $j$ of the encoder can be formulated as:
\begin{equation}
    Z_{\text{Trans}}^{(j)} = f_{\text{TransBlock}}^{(j)} \left( f_{\text{ResidualEnc}}^{(j)}(Z^{(j-1)}) \right)
\end{equation}
Furthermore, we propose a multi-scale fusion layer that aggregates the representations from all Transformer stages and the bottleneck latent space. This layer facilitates a global exchange of information across different resolutions. The fused features are subsequently projected back to the dimensions of the original latent space. The fusion operation  $f_{\text{Fusion}}$ is defined as:
\begin{equation}
    D^{(J)} = f_{\text{Fusion}} \left( [ Z^{(J)} , Z^{(1)}_{\text{Trans}}, Z^{(2)}_{\text{Trans}}, \dots, Z^{(J)}_{\text{Trans}}] \right)
\end{equation}
where $[\cdot]$ denotes channel-wise concatenation. Furthermore, we propose that the decoder maintains the traditional skip connections from each residual SE encoder stage, as defined in Equation (\ref{equ_decoder}), and utilizes the output of the fusion convolutional layer $D^{(J)}$ rather than the direct features of transformer blocks. This design prevents a significant increase in parameter count while enabling the model to simultaneously leverage multi-scale features from the residual blocks. In addition, the decoder inherits the global context captured by the Transformer blocks via the fused bottleneck features. Given that radiation dose maps are inherently non-negative, a Softplus activation layer is implemented at the output of the network to ensure this consistency and to avoid vanishing gradients for large positive values. \\
Given a training dataset of $N$ pair training data $\mathbb{D}_{\text{train}} = \left\{ \ith{V_{CT}}, \ith{X}, \ith{\alpha}, \ith{Y},  \right\}_{i=1}^N$, data-driven learning methods aim to minimize the error between the estimated doses and the ground truth (GT) doses as :

\begin{equation}
    \min_{\theta} \mathcal{L} \left( f_\theta \left( \ith{V_{CT}}, \ith{X}, \ith{\alpha} \right) - \ith{Y} \right)
    \label{eq:argmin_L}
\end{equation}

where $\mathcal{L}$ represents a loss function such as the $\ell_2$ distance (Mean Squared Error MSE) such as in \cite{javaid2019mitigating} or $\ell_1$ distance (Mean Absolute Error MAE) \cite{zhang2023deep}. Another approach is to use a combined loss function such as gradient-based MAE \cite{kraus2026single} or Gamma Index-based loss functions \cite{martinot2023differentiable}. \\

A significant challenge in medical imaging datasets is the high variability of input matrix dimensions between different subjects. Standard preprocessing techniques, such as global resizing, often introduce non-physical deformations to anatomical textures, potentially leading to incorrect feature representations. Cropping one image to a fixed dimension restricts usable regions of interest and limits the training sample size. To overcome these limitations, we propose to use a patch-based learning strategy maintained at a standardized isotropic resolution (e.g., $2 \times 2 \times 2 \text{ mm}^3 $). This approach is particularly advantageous where the sample size of datasets are limited, as it increases the effective training size and statistical diversity, significantly mitigating the risk of overfitting. Furthermore, this strategy preserves local spatial information, allowing the network to learn invariant physical features of radiation transport on dose maps and anatomical textures of CT scans. By training on local patches, the model achieves superior convergence and generalizes more effectively to the various anatomical scales encountered in different training sets.

\section{Results}
\begin{figure*}[t] 
    \begin{minipage}[b]{1\linewidth}
    \centering
    \includegraphics[width=1\linewidth]{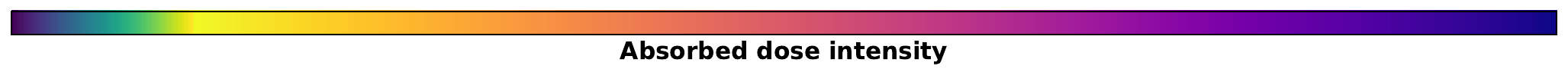} 
    \end{minipage} 
    
  \tabcolsep=0pt
  \begin{tabular*}{\textwidth}{@{\extracolsep{\fill}}cccc}
    \includegraphics[width=0.24\linewidth]{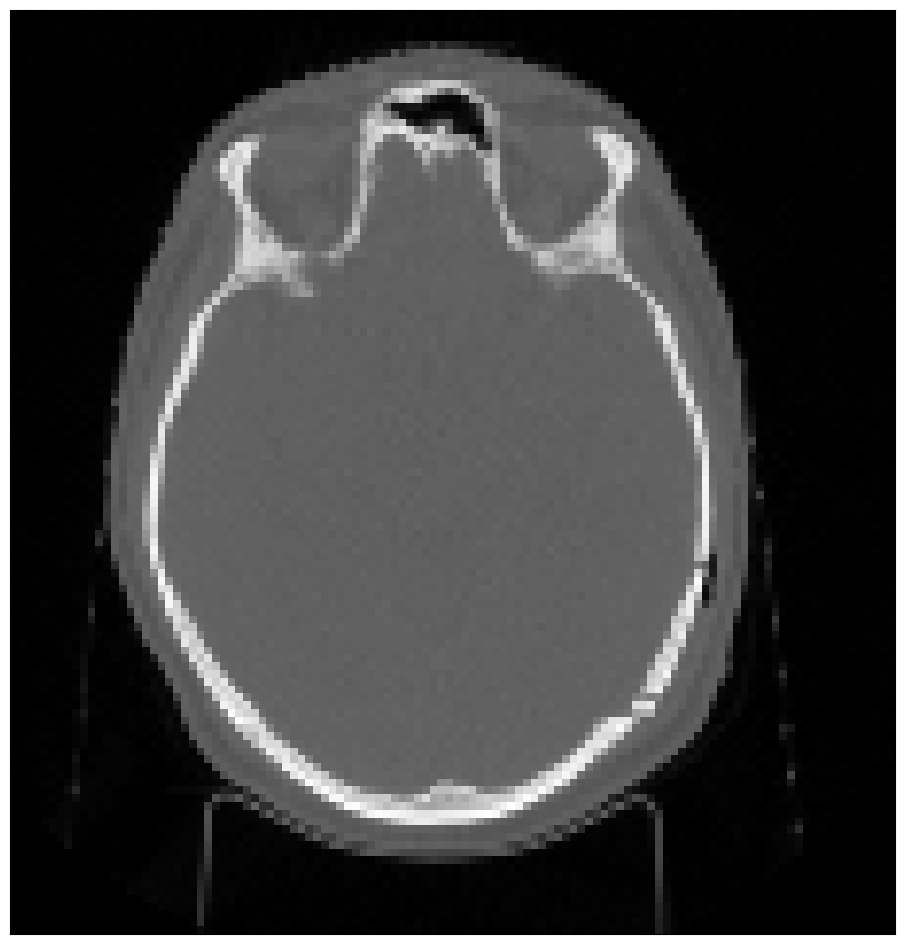}&
    \includegraphics[width=0.24\linewidth]{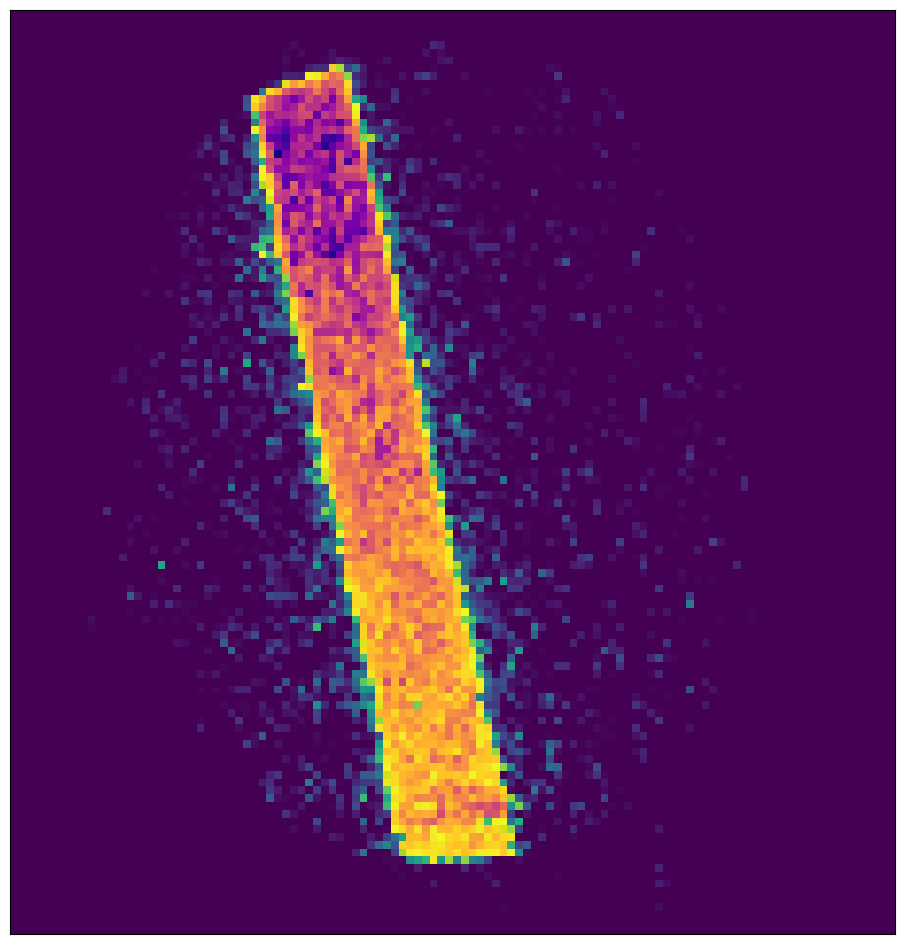}  &
    \includegraphics[width=0.24\linewidth]{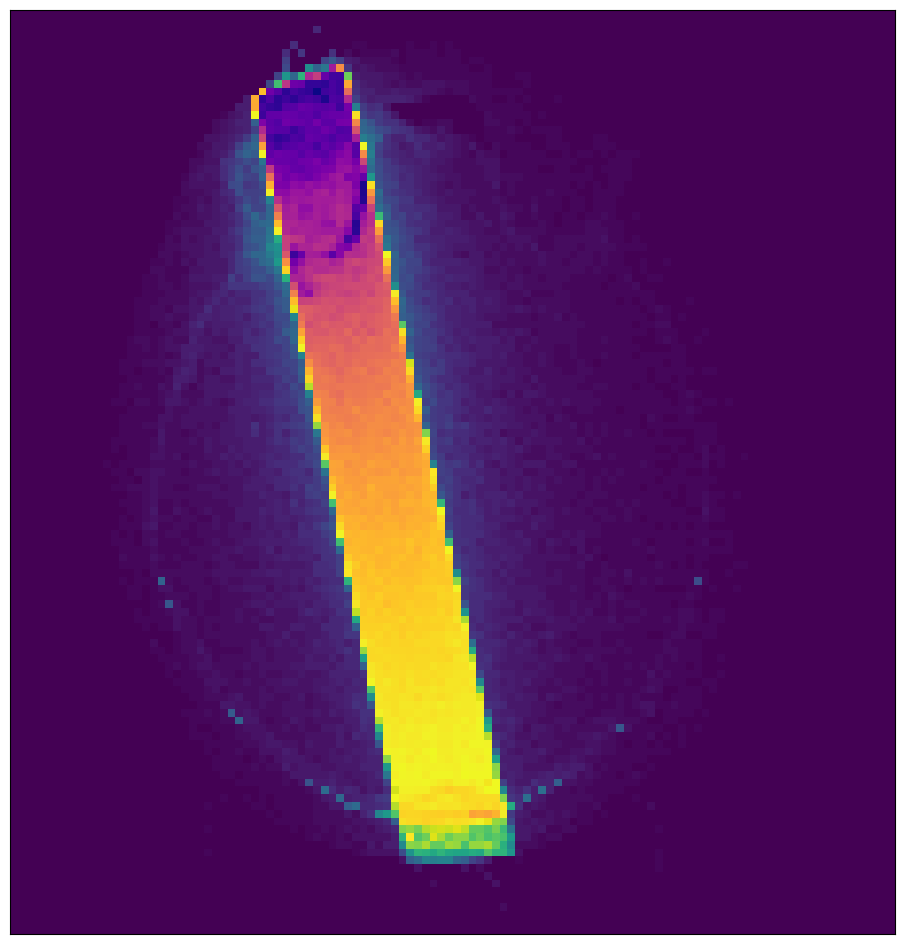}  &
    \includegraphics[width=0.24\linewidth]{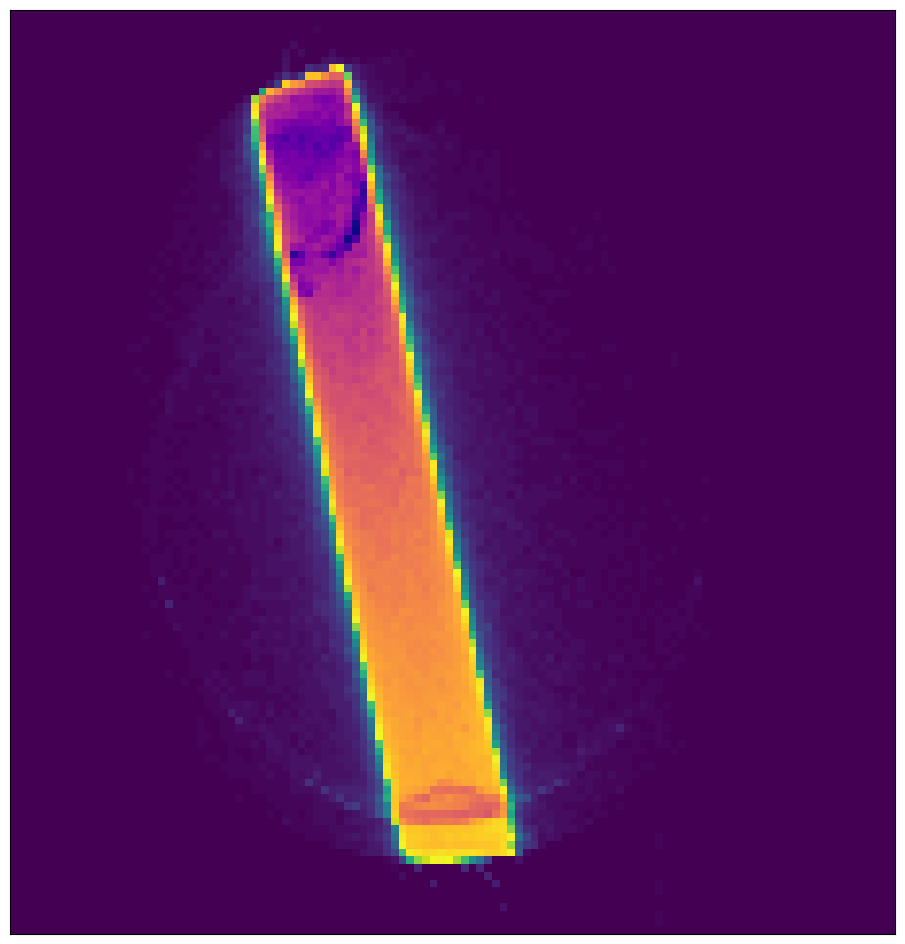}  \\
    \small (a) CT scan &
    \small (b) LS TrueBeam input &
    \small (c) Mono-energetic input &
    \small (d) TrueBeam reference \\

    \includegraphics[width=0.24\linewidth]{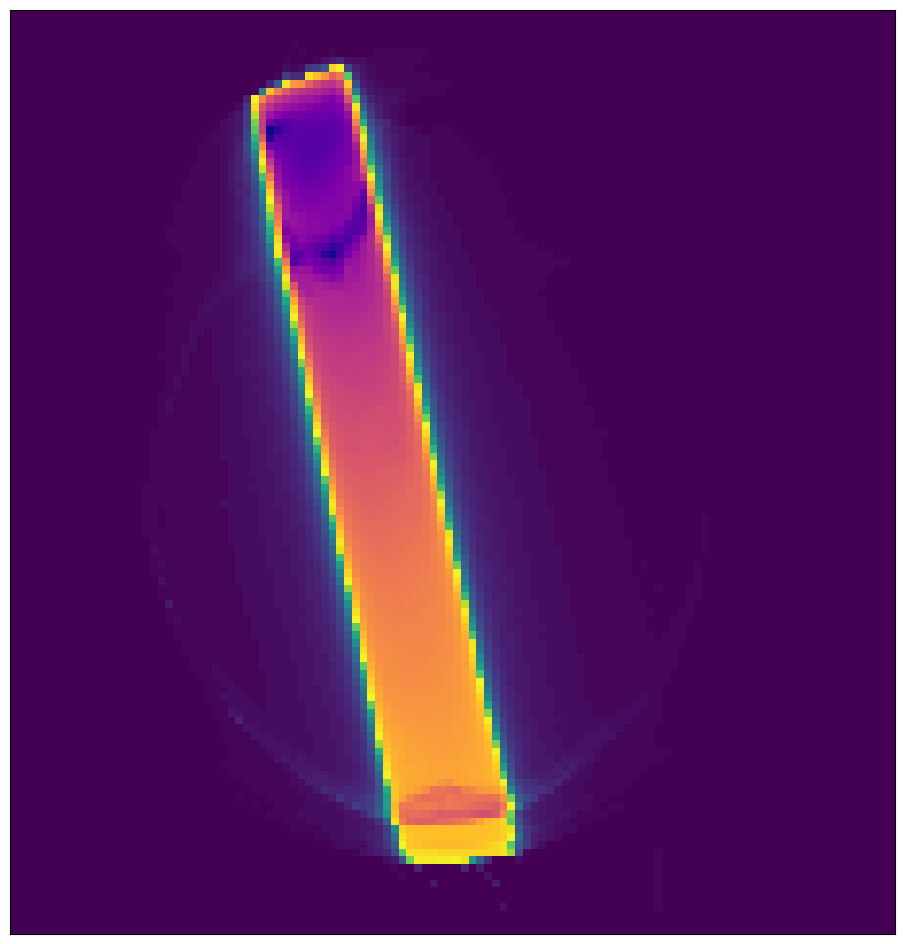} &
    \includegraphics[width=0.24\linewidth]{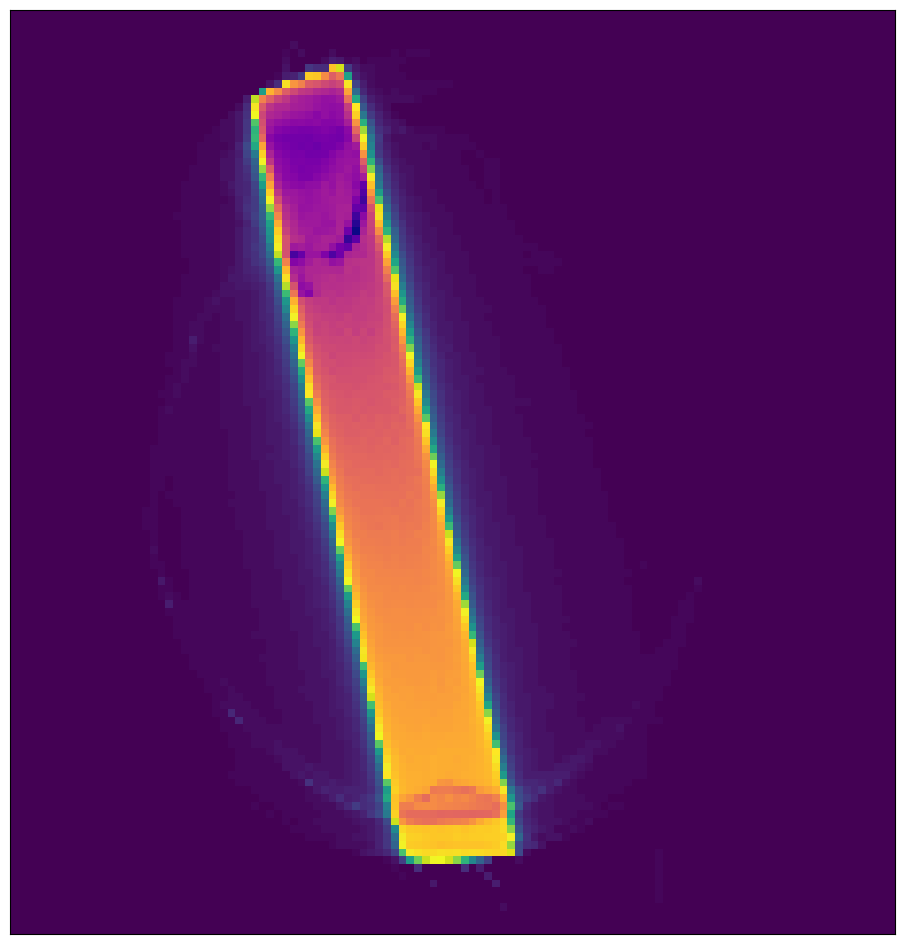}  &
    \includegraphics[width=0.24\linewidth]{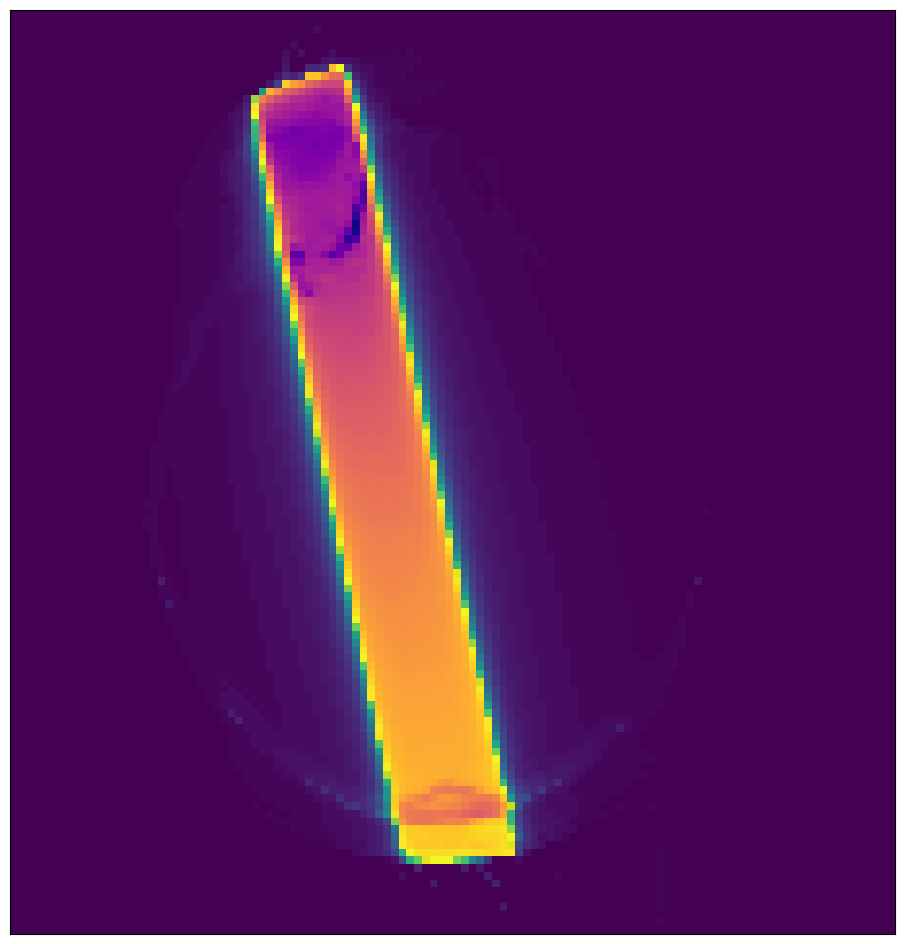} &
    \includegraphics[width=0.24\linewidth]{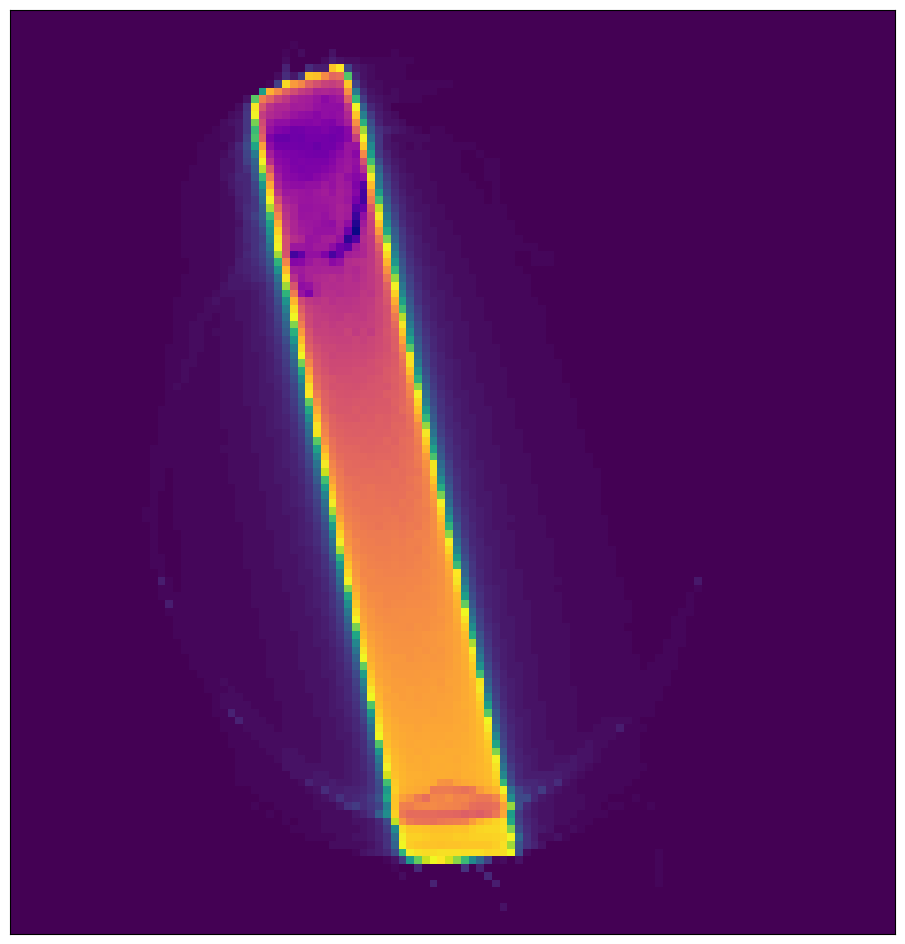}  \\
    \includegraphics[width=0.24\linewidth]{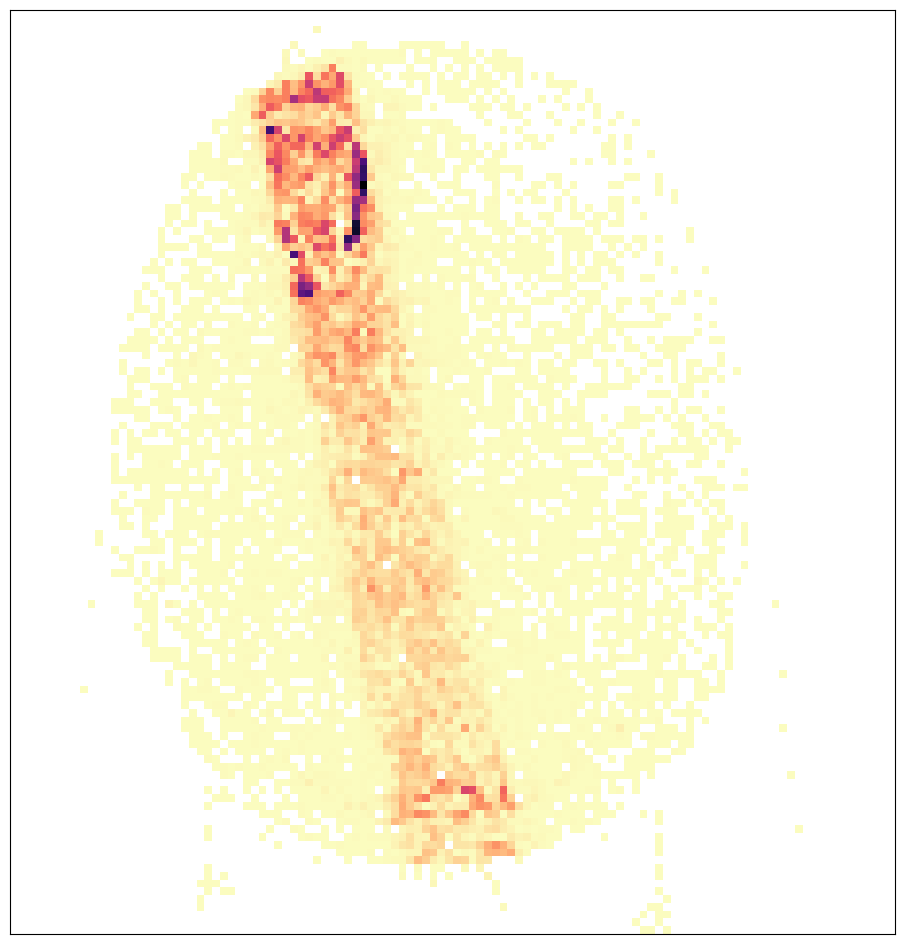} &
    \includegraphics[width=0.24\linewidth]{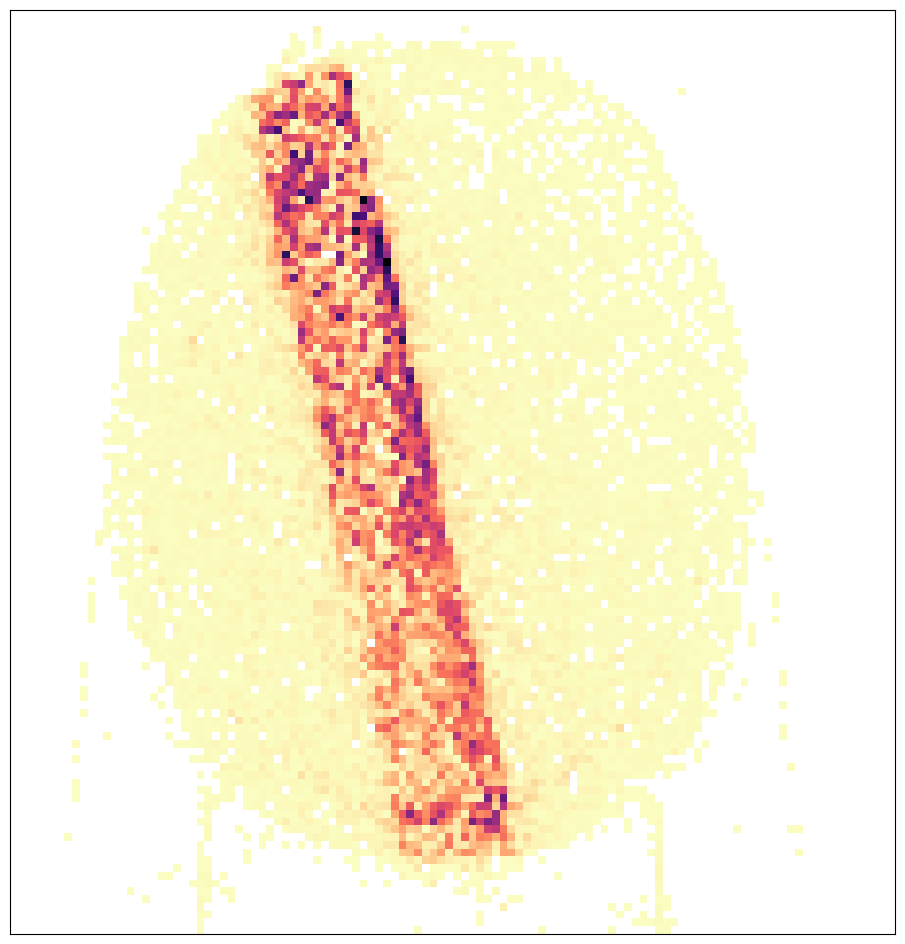} &
    \includegraphics[width=0.24\linewidth]{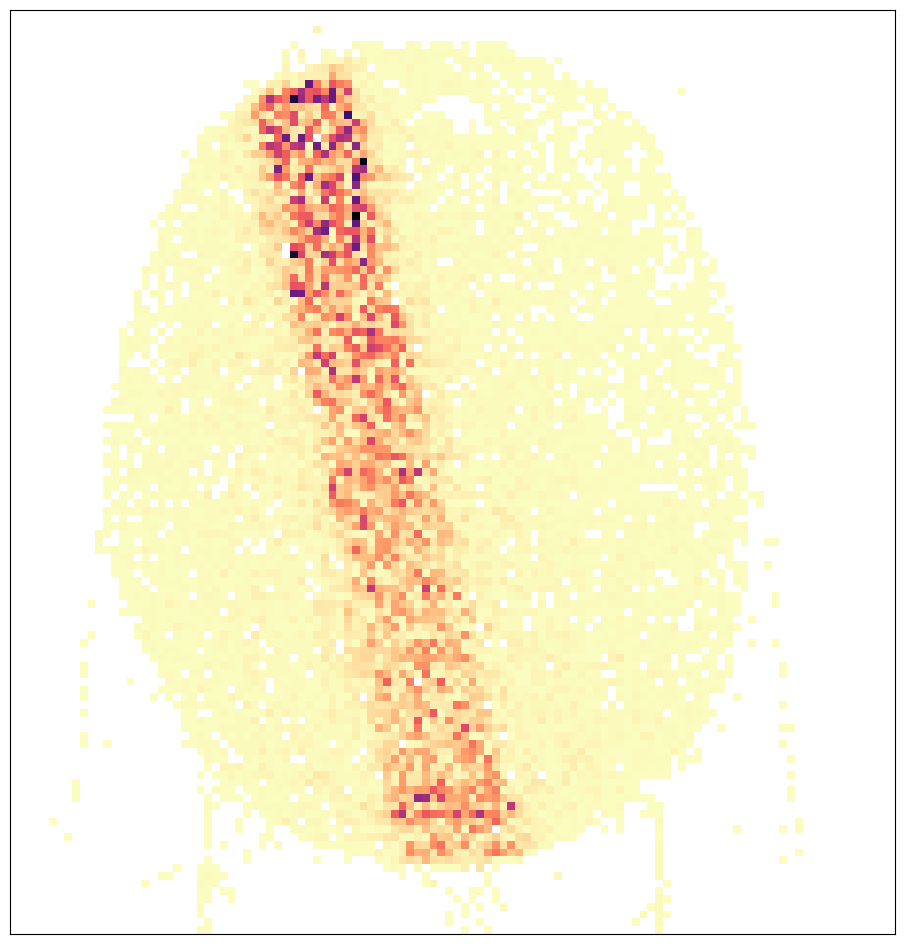} &
    \includegraphics[width=0.24\linewidth]{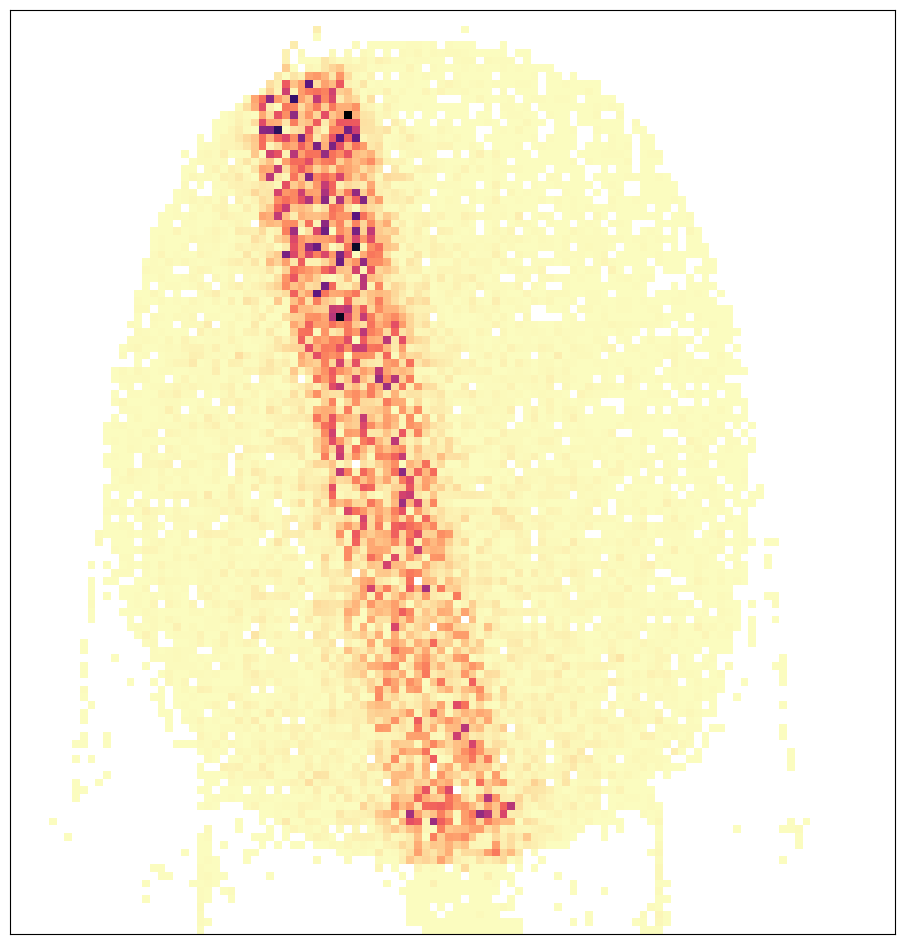} 
    \\
    \small (e) Denoising of LS input&
    \small (f) Denoising with CT &
    \small (g) Energy-Shifting &
    \small (h) Energy-Shifting with CT \\
  \end{tabular*}
  
    \begin{minipage}[b]{1\linewidth}
    \centering
    \includegraphics[width=1\linewidth]{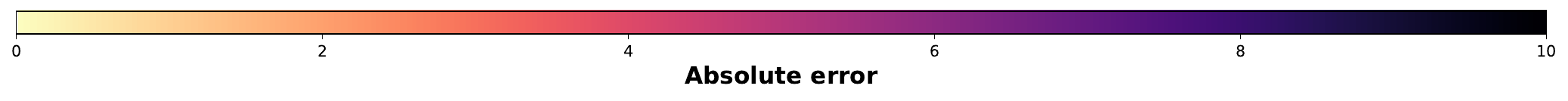} 
    \end{minipage} 

  \caption{Illustration of an axial slice predicted using the same training/testing protocol by different learning approaches : denoising and Energy-Shifting trained and evaluated on Head dataset. The first row shows CT scan, the input of two approaches and the reference dose. The second row shows the prediction of each approach without/with CT scan. The last row shows the absolute difference between the predictions and the reference.}
  \label{results_head_gamma_0948} 
\end{figure*}

\subsection{Experimental setup}

The proposed models were implemented using the PyTorch framework. The datasets Head and Pelvis were initially partitioned into training and testing sets using a 9:1 ratio. Subsequently, the training set was further subdivided into training and validation subsets using an 8:2 ratio.  No data normalization was applied to the 3D images (CT scans and dose maps) to preserve the original physical intensities. The beam parameters are normalized using $\ell_2$ normalization. All networks were trained for 2000 epochs with a mini-batch size of 4. During training, input patches of $128 \times 128 \times 64$ voxels were sampled via random cropping from the original 3D volumes. We utilized the AdamW optimizer with an initial learning rate of $10^{-4}$, and the \textit{ReduceLROnPlateau} scheduler was employed to dynamically adjust the learning rate based on validation loss. During the inference phase, a sliding-window approach, as implemented by MONAI platform \cite{monai2020project}, was used with an overlap of 0.9 to apply the trained models to whole 3D testing images. To minimize boundary artifacts, a Gaussian weighting function was applied to the overlapping regions, assigning lower weight to predictions at the patch edges. Model performance was evaluated on the testing set using Peak Signal-to-Noise Ratio (PSNR) and Gamma Passing Rate (GPR). The gamma index was configured at 3\%/3mm with a 10\% low-dose threshold using Gate Tools \cite{sarrut_gate_2025}, consistent with previous studies \cite{bresciani2013tomotherapy,song2015gamma,zhang2023deep}. The segmentation map of the testing images is generated by using TotalSegmentator tool \cite{wasserthal2023totalsegmentator}.

\subsection{Results of the proposed Energy-Shifting}

\begin{table}[ht]
 \centering
 \begin{tabular}{l|l|l|c}
  \toprule
    Learning & Training set : Head & \multicolumn{2}{c}{Testing set}  \\
    \cline{2-4}
    approachs & Input types &  Head & Pelvis \\ 
    \midrule
    \multirow{2}{*}{Denoising}  & Low-sampling TrueBeam & 52.83  &  49.64 \\
    \cline{2-4}
                                & Low-sampling TrueBeam, CT & 56.43  & 46.33 \\
    \hline
    \multirow{2}{*}{Energy-Shifting} & Mono beam & 56.68  &  48.63 \\
    \cline{2-4} 
                            & Mono beam, CT & \textbf{57.84} & \textbf{50.51} \\
    \bottomrule
  \end{tabular}
  \caption{Quantitative evaluation using PSNR ($\text{dB}$) of UNet3D-based learning methods trained on dataset Head and then tested on two datasets Head and Pelvis with respect to different input types.}
  \label{tab:ablation_study_input}
\end{table}

We evaluated the performance of the proposed Energy-Shifting-based dose calculation against a standard denoising approach using a 3D UNet baseline. For the denoising task, the input consisted of a low-statistic dose map ($2 \times 10^5$ particles) targeted at predicting a high-statistic distribution of $6$ MV TrueBeam ($2 \times 10^7$ particles). In contrast, the proposed energy-shifting approach utilized a 500 keV mono-energetic dose map with $2 \times 10^7$ particles as the input to predict the same TrueBeam output. The models were trained with the same settings such as the MAE loss function and the optimization method and tested on Head dataset. To assess the generalization of two approaches, all models were also evaluated on an unseen Pelvis dataset. As shown in Table \ref{tab:ablation_study_input}, the concatenation of CT volumes as a second input channel significantly improved results across all methods within the same set by providing essential anatomical context. However, CT volumes did not improve the results of the denoising approach when testing on Pelvis dataset. Furthermore, the Energy-Shifting strategy consistently outperformed the denoising approach within Head testing set. The results of the Energy-Shifting approach on Pelvis dataset indicate that the proposed learning method maintains superior robustness and accuracy on unseen anatomy. 

Figure  \ref{results_head_gamma_0948} illustrates two fast Monte Carlo-based dose calculation methods using UNet architecture: standard denoising and our proposed Energy-Shifting approach. Each method was tested using both a single-input configuration and a CT-concatenated input strategy. To facilitate the evaluation of the models, results are presented as absolute difference maps relative to the ground-truth Monte Carlo simulation. We observe that the integration of CT data provides critical structural guidance, allowing the model to preserve complex anatomical features in the denoising baseline, such as the orbital bone proximal to the source. Furthermore, the Energy-Shifting method significantly preserves the texture of the beam border. In addition, the use of an element-wise loss function results in a smoothing effect compared to the original simulated reference. However, this effectively suppresses stochastic Monte Carlo noise.

Finally, we investigated the impact of the loss function by comparing Mean Squared Error (MSE) and Mean Absolute Error (MAE). Comparison of loss functions in Table \ref{tab:ablation_study_loss} reveals that MAE outperformed MSE, providing a more accurate reconstruction of the dose distribution. This suggests that MAE is more robust to outliers, providing a higher-fidelity reconstruction of the dose maps. \\

\begin{figure*}[t] 
  \begin{minipage}[b]{0.24\linewidth}
    \centering
    \includegraphics[width=1\linewidth]{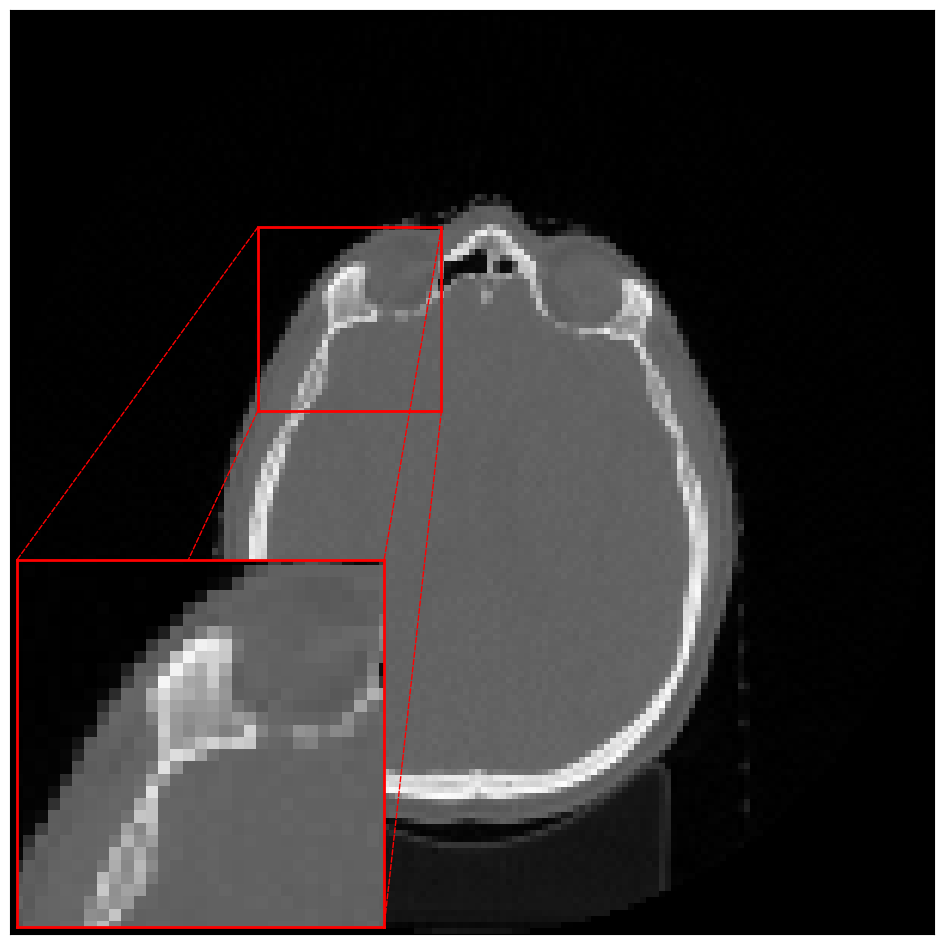} 
    \caption*{CT scan} 
  \end{minipage} 
  \begin{minipage}[b]{0.24\linewidth}
    \centering
    \includegraphics[width=1\linewidth]{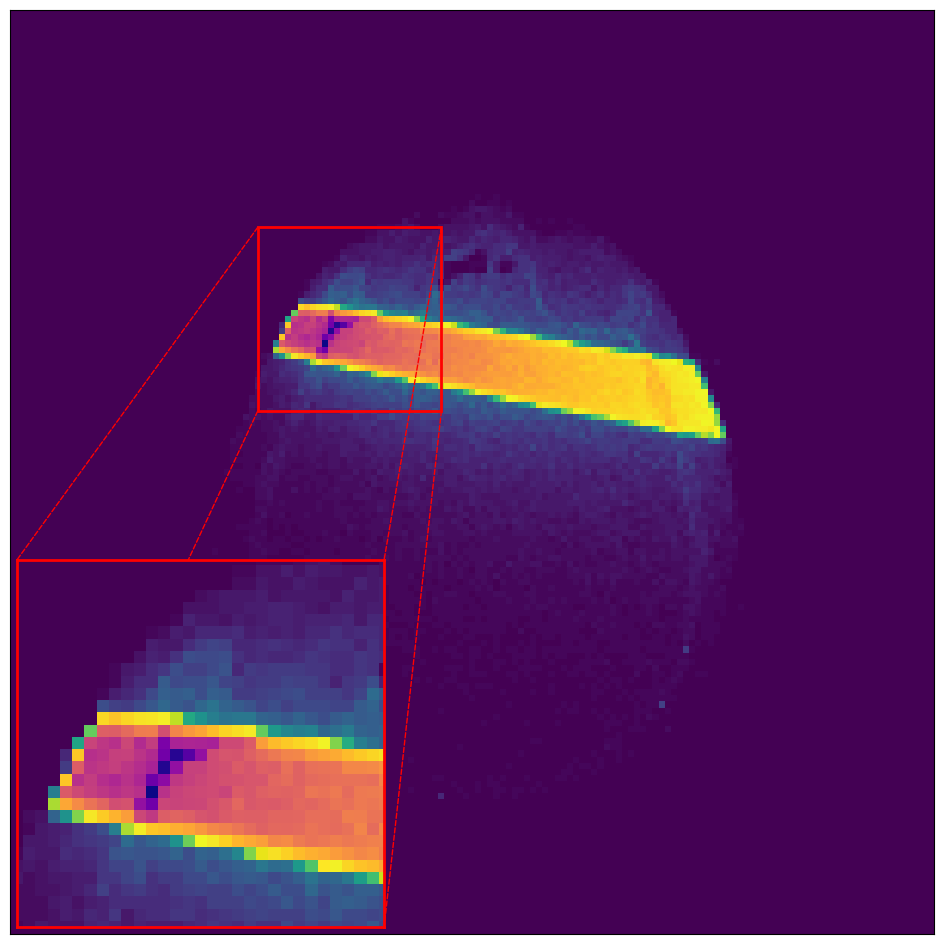} 
    \caption*{Mono-energetic input} 
  \end{minipage}
  \begin{minipage}[b]{0.24\linewidth}
    \centering
    \includegraphics[width=1\linewidth]{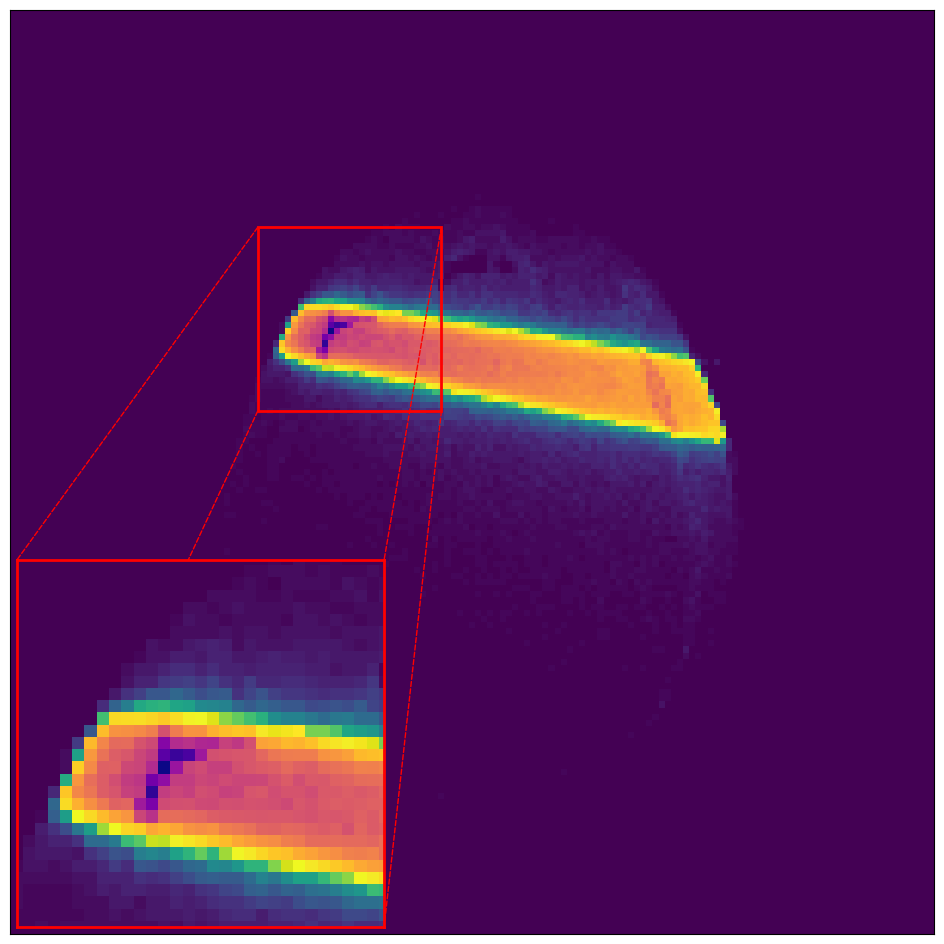} 
    \caption*{TrueBeam reference} 
  \end{minipage} 
  \begin{minipage}[b]{0.24\linewidth}
    \centering
    \includegraphics[width=1\linewidth]{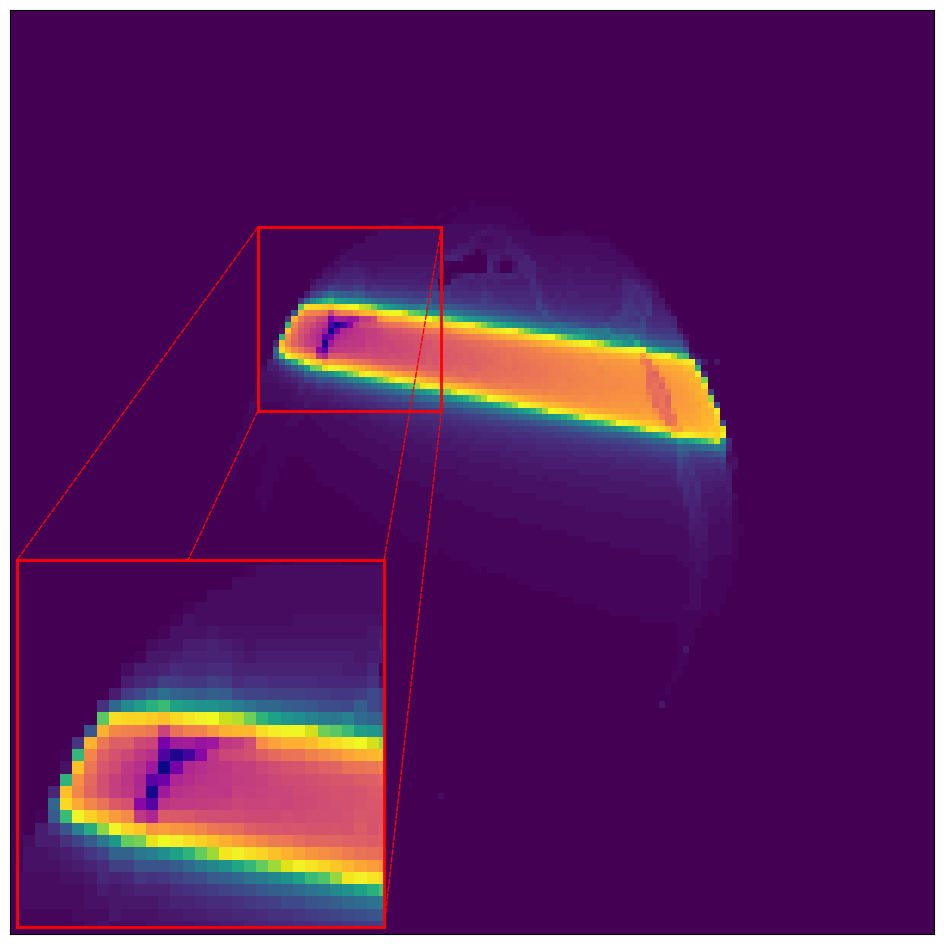} 
    \caption*{UNet3D} 
  \end{minipage} 
  \hfill
  \vspace{5pt}
  \begin{minipage}[b]{0.24\linewidth}
    \centering
    \includegraphics[width=1\linewidth]{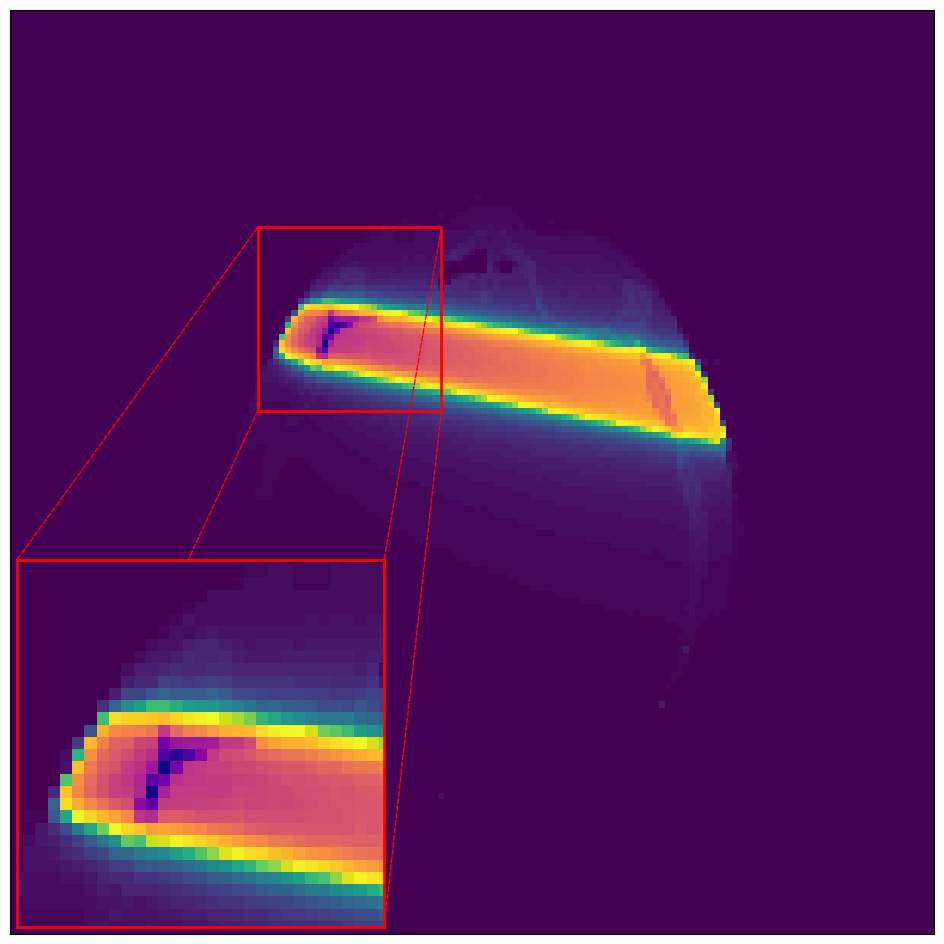} 
    \caption*{ResidualUNet3D} 
  \end{minipage} 
  \begin{minipage}[b]{0.24\linewidth}
    \centering
    \includegraphics[width=1\linewidth]{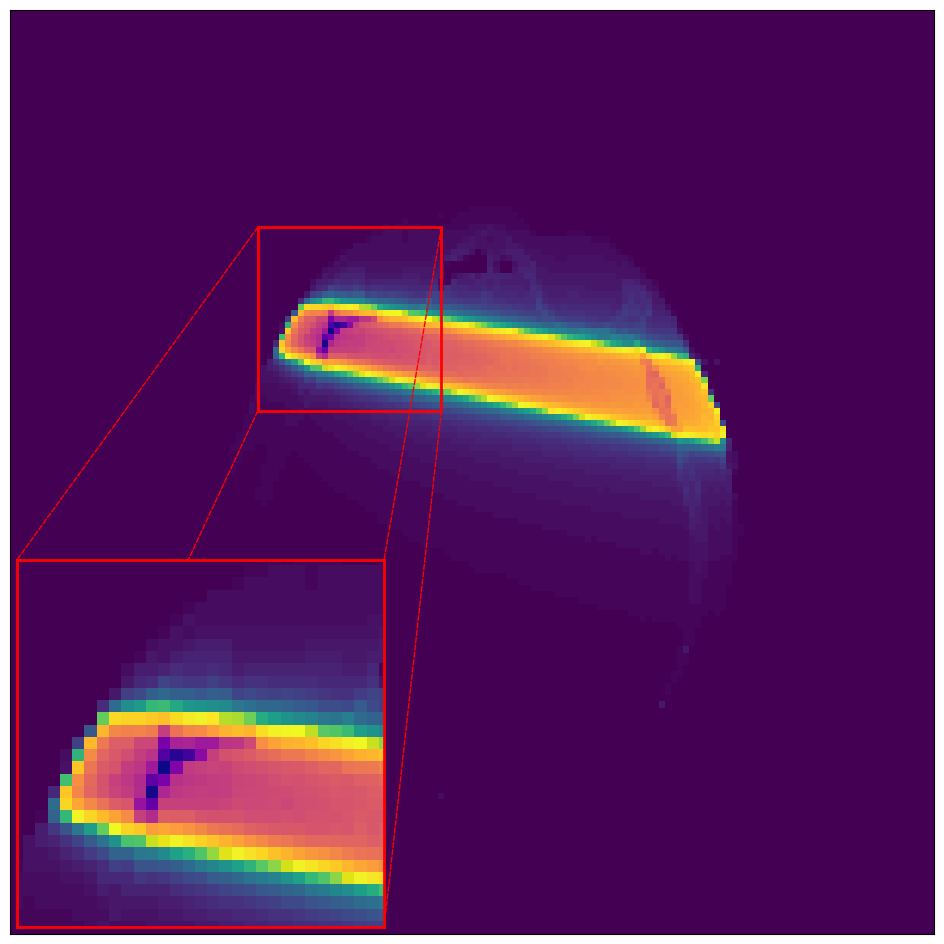} 
    \caption*{UNETR} 
  \end{minipage} 
  \begin{minipage}[b]{0.24\linewidth}
    \centering
    \includegraphics[width=1\linewidth]{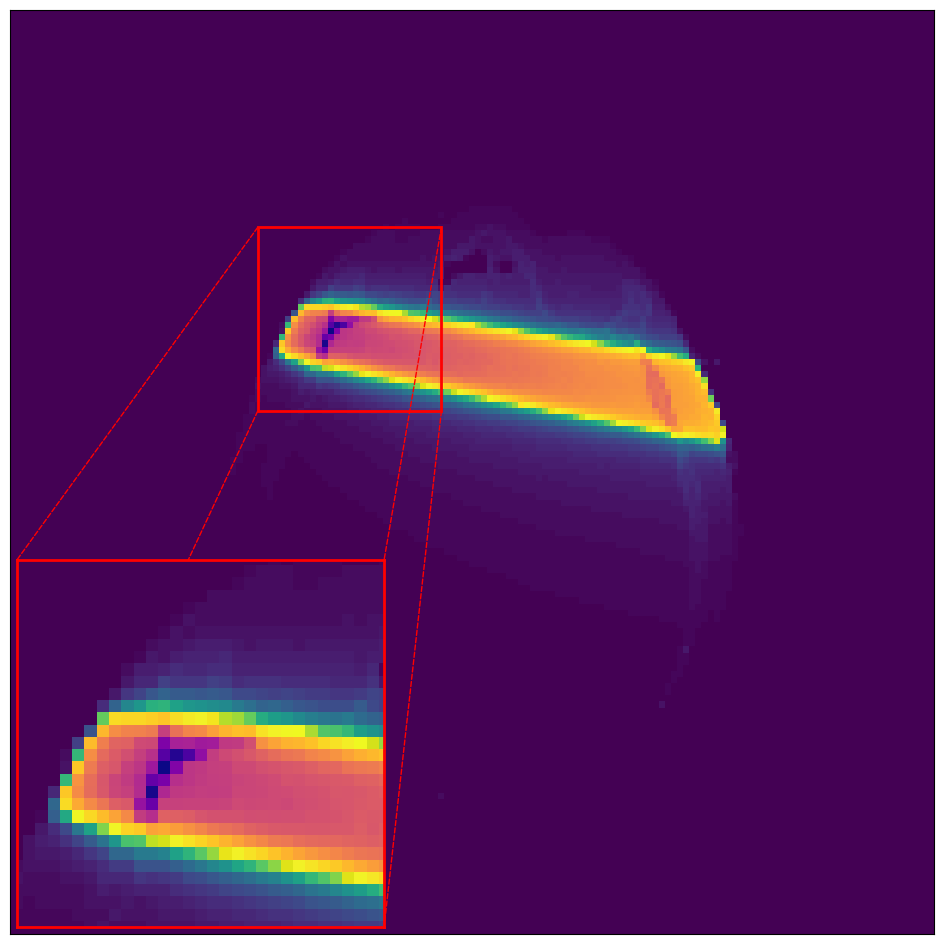} 
    \caption*{SwinUNETR} 
  \end{minipage} 
  \begin{minipage}[b]{0.24\linewidth}
    \centering
    \includegraphics[width=1\linewidth]{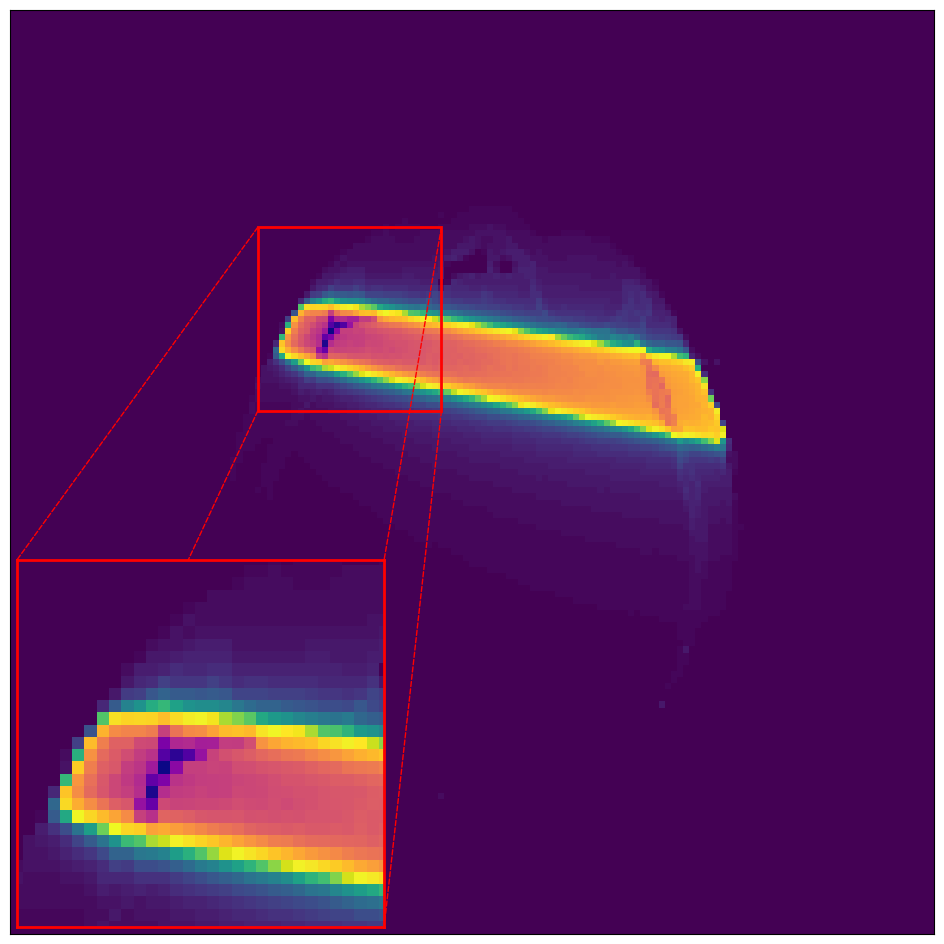} 
    \caption*{Proposed TransUNetSE3D} 
  \end{minipage} 
  \\
    \begin{minipage}[b]{1\linewidth}
    \centering
    \includegraphics[width=0.95\linewidth]{results/colorbar.pdf} 
    \end{minipage} 
  \caption{Illustration of an axial slice predicted by different architectures using UNet3D, ResidualUNet3D, UNETR, SwinUNETR and TransUNetSE3D trained and tested on Head dataset. We show the zoom regions in the slices at the left corner by a red border.}
  \label{results_head_gamma500keV_0984_slice80} 
\end{figure*}

\begin{table}[t]
 \centering
 \begin{tabular}{l|l|c}
  \toprule
   Training set : Head & \multicolumn{2}{c}{Testing set}  \\
    \hline
    Loss function  &  Head & Pelvis \\ 
    \midrule
    Mean squared error (MSE) &  57.44  &  48.71 \\  
    Mean absolute error (MAE) &  \textbf{57.84}  &  \textbf{50.51}  \\
    \bottomrule
  \end{tabular}
  \caption{Quantitative evaluation using PSNR ($\text{dB}$) of UNet3D-based Energy-Shifting learning methods trained on dataset Head and then tested on two datasets Head and Pelvis with respect to different loss functions. The best results are highlighted in bold. }
  \label{tab:ablation_study_loss}
\end{table}

\subsection{Results of the proposed architecture}

We evaluated the Energy-Shifting approach across several architectures, including UNet3D, ResidualUNet3D, UNETR, SwinUNETR, and our proposed TransUNetSE3D model. The results are summarized in Table \ref{tab:comparison_architectures}. The performance of ResidualUNet3D significantly exceeds those of the standard UNet3D due to the integration of residual blocks within the hidden convolutional blocks. Indeed, ResidualUNet3D aggregates features from the encoder stages and the upsampled decoder outputs via element-wise summation rather than the direct mapping of plain UNet3D using channel-wise concatenation. This architectural choice is based on the principle that it is easier for a network to learn a residual mapping, essentially a small adjustment of features, than to learn an entirely new map from scratch. Furthermore, within the same anatomical Head datasets, Transformer-based models such as UNETR and SwinUNETR outperformed standard convolutional baselines (\textit{e.g.} UNet3D and ResidualUNet3D). However, these two transformer-based models suffer from significant performance degradation when evaluated on the unseen Pelvis dataset in terms of PNSR and GPR. In contrast, ResidualUNet3D demonstrated robust generalization across both different anatomical testing sets. Our proposed TransUNetSE3D model effectively mitigates the degradation observed in other transformer-based architectures, maintaining high performance and robustness on both Head and Pelvis sets by exploiting the local stability of residual blocks. In addition, leveraging the global context of patches captured by transformer blocks, our proposed architecture outperforms both classic CNNs and recent transformer-based models with regard to PSNR and GPR.

Figure \ref{results_head_gamma500keV_0984_slice80} illustrates the qualitative performance of the compared architectures when trained and evaluated within the same anatomical domain. As illustrated, the beam profile of the mono-energetic input differs significantly from the reference TrueBeam dose distribution. The primary objective of the compared models is to transform the simplified mono-energetic beam shape into the complex poly-energetic reference. This strategy is motivated by computational efficiency: mono-energetic dose maps can be generated instantaneously using the GPU-based MC simulation such as GGEMS \cite{bert_geant4-based_2013}, while producing the reference dose based on LINAC via MC simulations (such as OpenGATE) is a time-consuming process, often requiring several hours of computation. In this intra-domain setting (Head dataset), all models exhibit high fidelity to the reference dose, accurately preserving both the beam profile and the spatial dose distribution. This qualitative agreement is supported by quantitative results, with all methods achieving a GPR larger than $99\%$. However, while performance is consistent on the same anatomy, a central objective of this work is to analyze the robustness and generalizability of these architectures when applied to unseen anatomical structures.

\begin{figure*}[ht] 
  \begin{minipage}[b]{0.245\linewidth}
    \centering
    \includegraphics[width=1\linewidth]{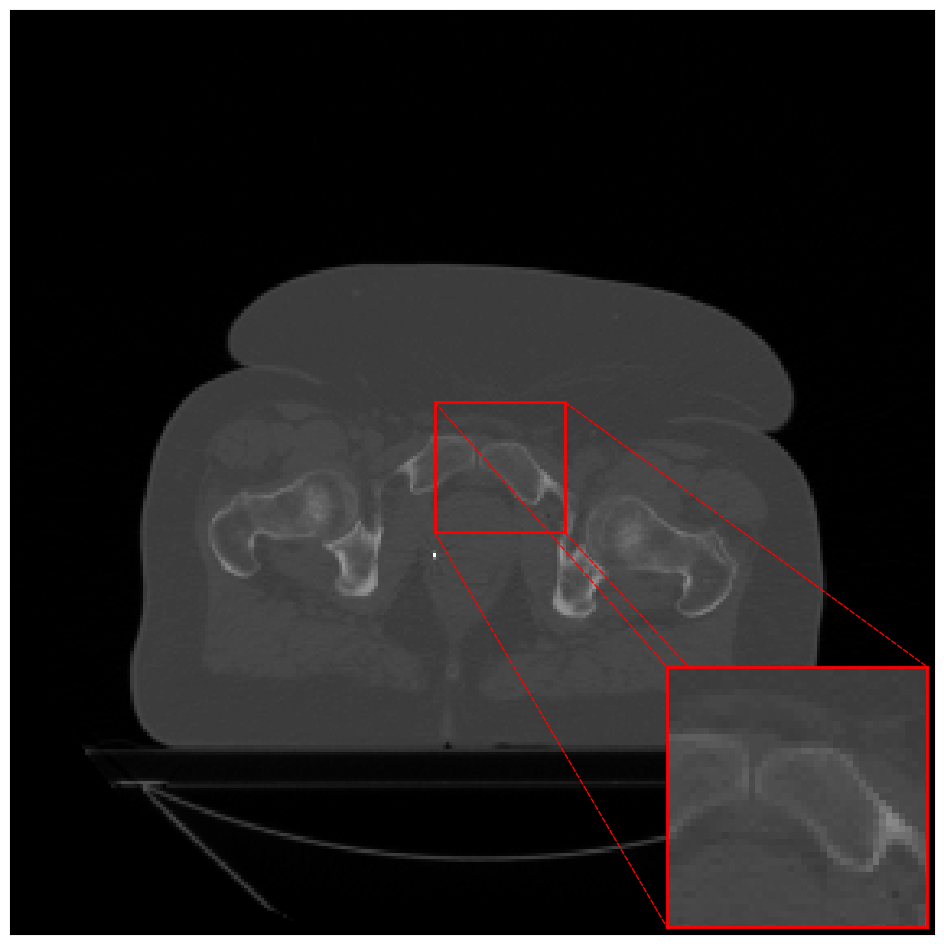} 
    \caption*{(a) CT scan} 
  \end{minipage} 
  \begin{minipage}[b]{0.245\linewidth}
    \centering
    \includegraphics[width=1\linewidth]{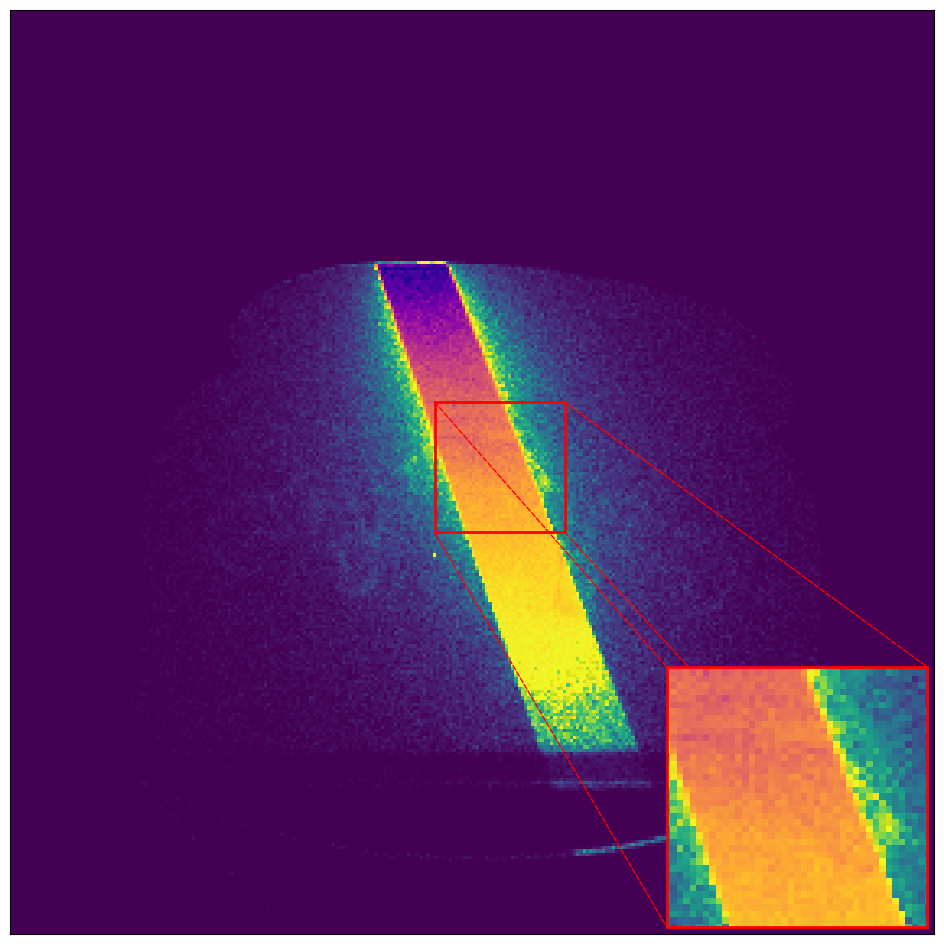} 
    \caption*{(b) Mono-energetic input} 
  \end{minipage}
  \begin{minipage}[b]{0.245\linewidth}
    \centering
    \includegraphics[width=1\linewidth]{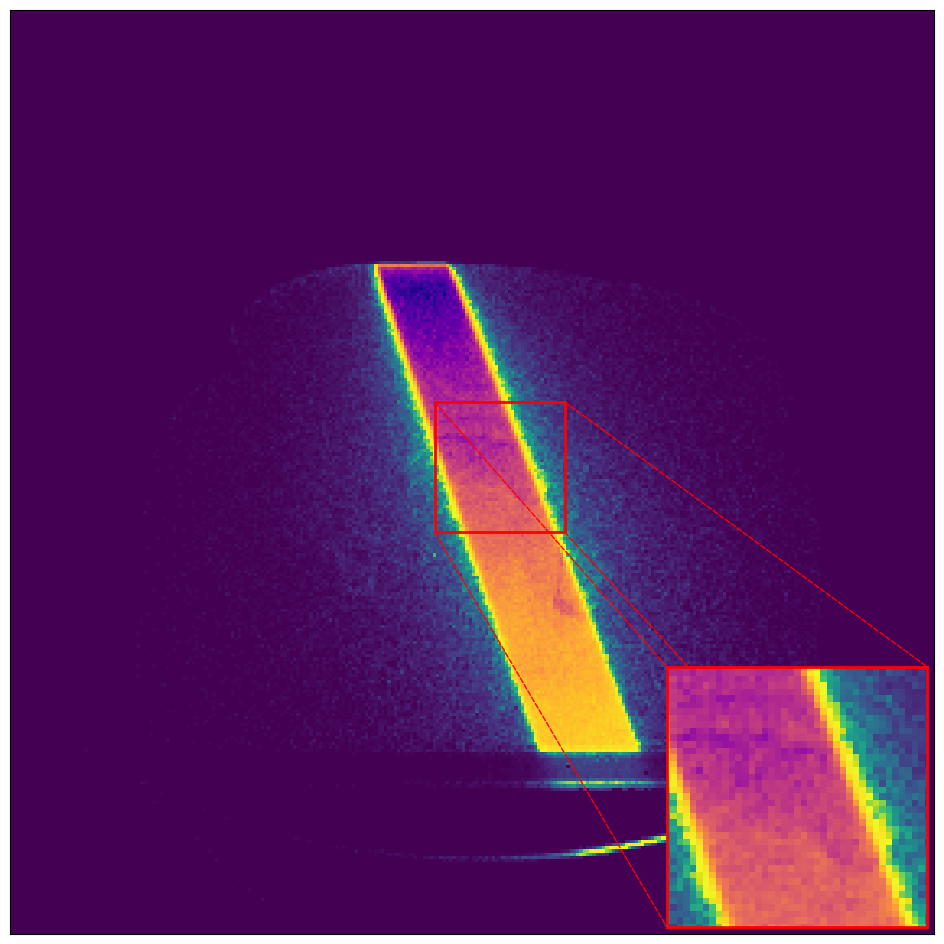} 
    \caption*{(c) TrueBeam reference} 
  \end{minipage} 
  \begin{minipage}[b]{0.245\linewidth}
    \centering
    \includegraphics[width=1\linewidth]{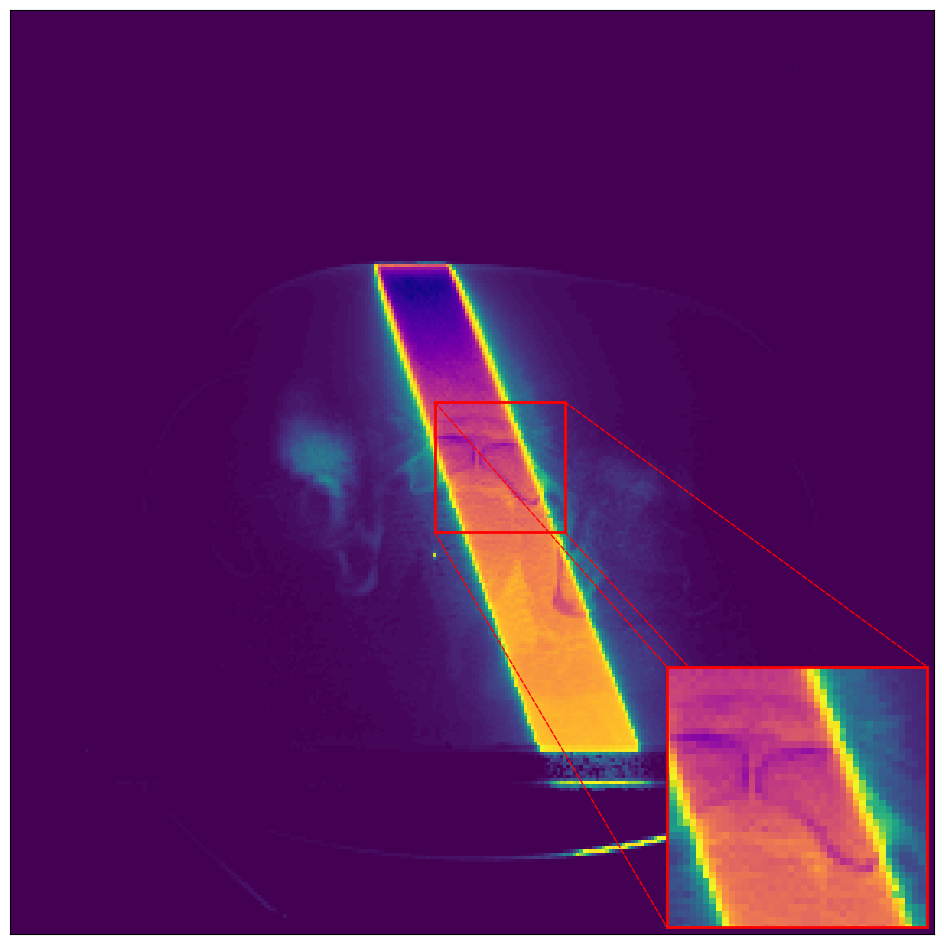} 
    \caption*{(d) UNet3D} 
  \end{minipage} 
  \hfill
  \begin{minipage}[b]{0.245\linewidth}
    \centering
    \includegraphics[width=1\linewidth]{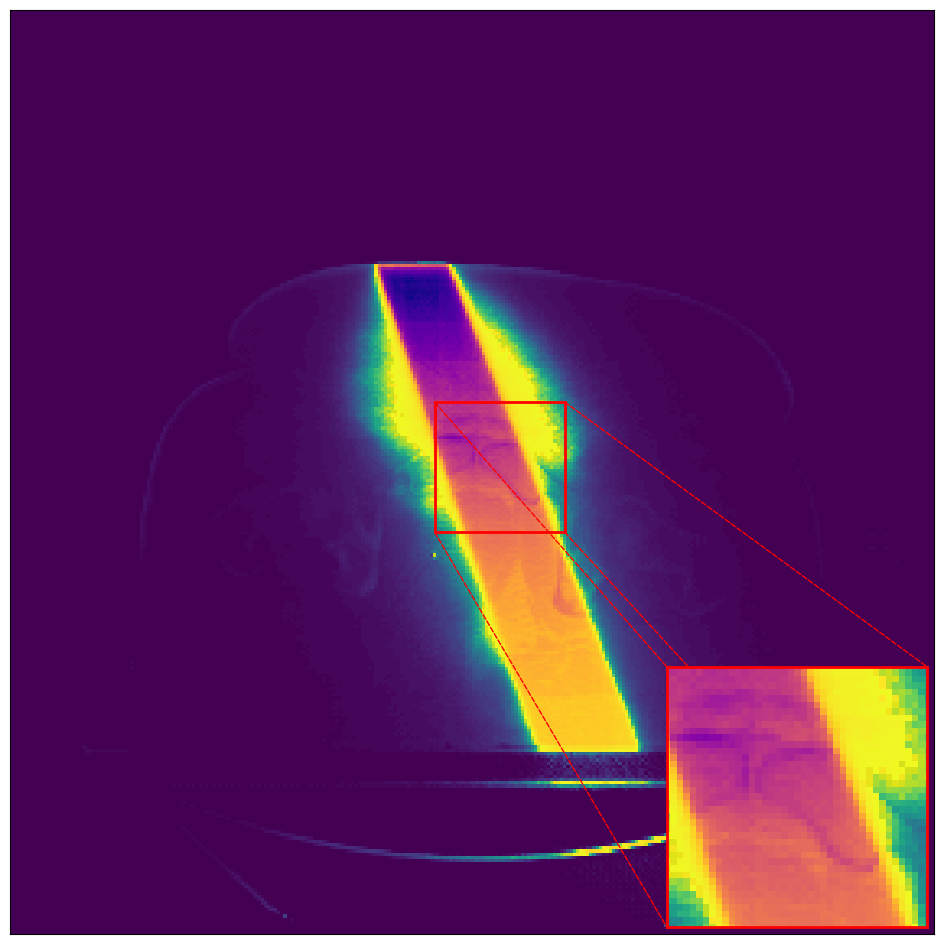} 
    \caption*{(e) ResidualUNet3D} 
  \end{minipage} 
  \begin{minipage}[b]{0.245\linewidth}
    \centering
    \includegraphics[width=1\linewidth]{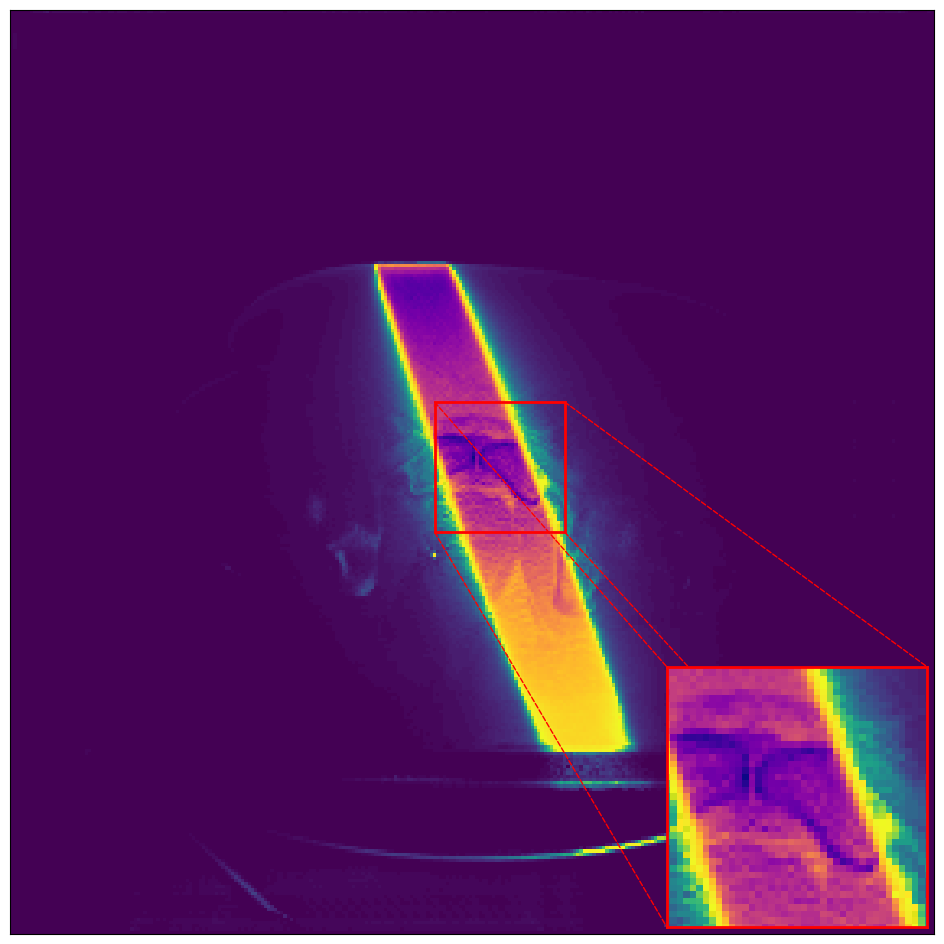} 
    \caption*{(f) UNETR} 
  \end{minipage} 
  \begin{minipage}[b]{0.245\linewidth}
    \centering
    \includegraphics[width=1\linewidth]{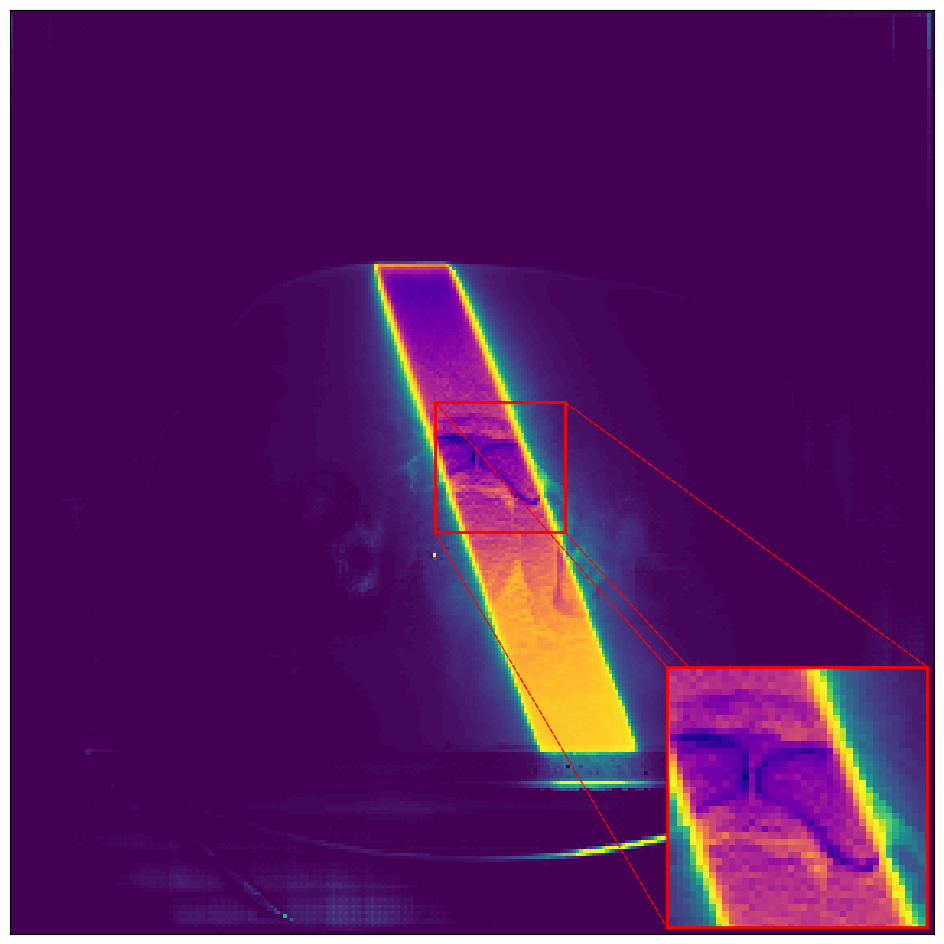} 
    \caption*{(g) SwinUNETR} 
  \end{minipage} 
  \begin{minipage}[b]{0.245\linewidth}
    \centering
    \includegraphics[width=1\linewidth]{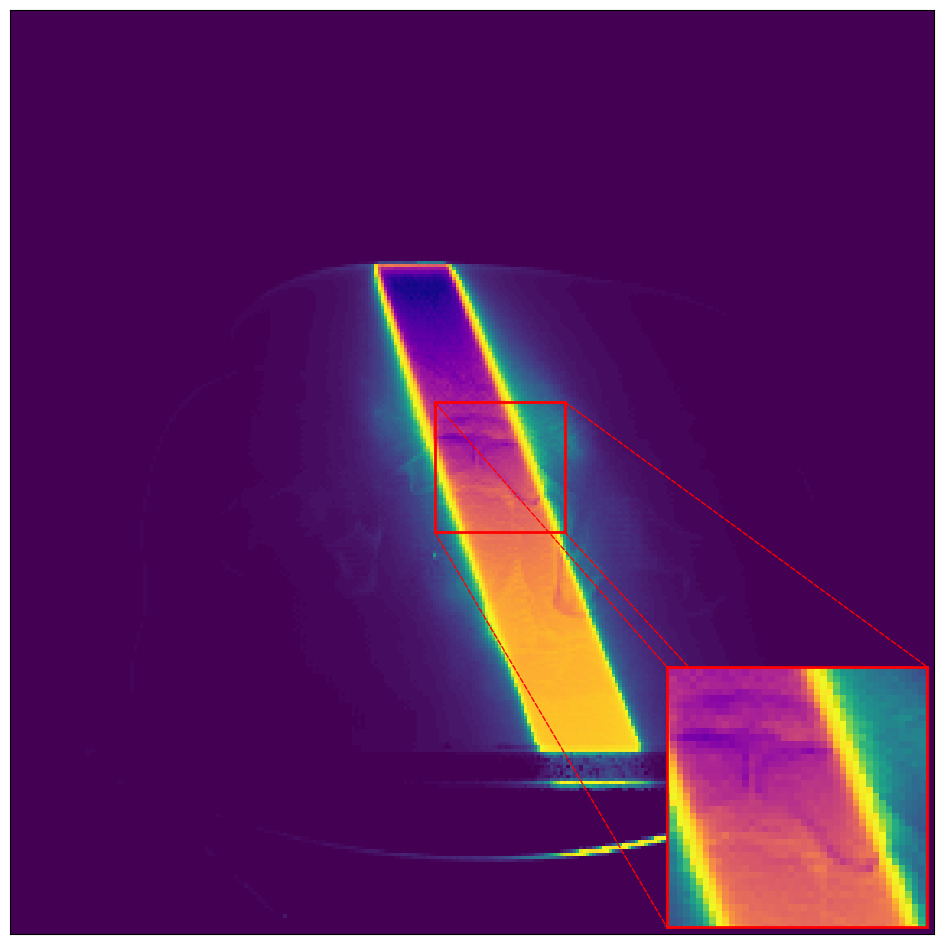} 
    \caption*{(h) TransUNetSE3D} 
  \end{minipage} 
  \hfill
  \begin{minipage}[b]{1\linewidth}
    \centering
    \includegraphics[width=0.98\linewidth]{results/colorbar.pdf} 
  \end{minipage}
  \caption{Illustration of an axial slice predicted by different architectures using the same training/testing protocol : UNet3D, ResidualUNet3D, UNETR, SwinUNETR and TransUNetSE3D trained on Head dataset and tested on Pelvis dataset. We show the zoom regions in the slices at the right corner by a red border.}
  \label{results_pelvis_gamma_0903} 
\end{figure*}

Figure \ref{results_pelvis_gamma_0903} illustrates the qualitative performance of the compared architectures, trained on the Head dataset and evaluated on the unseen Pelvis dataset. Notably, the two transformer-based models (\textit{i.e.} UNETR and SwinUNETR) exhibit significant structural artifacts at organ boundaries, leading to degraded predictions on out-of-distribution data. This suggests that while CT-guided features enhance representation within the same anatomy, they may not generalize effectively across different anatomical regions. In contrast, standard convolutional models such as UNET3D and ResidualUNet3D avoid these boundary distortions but fail to accurately preserve sharp dose gradients at the beam edges and around the beam contours. Our proposed architecture TransUnetSE3D effectively bridges this gap by leveraging the inherent stability of residual blocks along with the transformer’s ability to model long-range spatial correlations. As a result, the proposed model yields the most promising results, demonstrating high fidelity to the reference dose and superior robustness to unseen anatomy.

\begin{table}[ht]
    \centering
    \begin{tabular}{l|c|c|c|c}
    \hline
    Training set : Head & \multicolumn{4}{c}{Testing set}  \\
    \cline{1-5}
    Architectures & \multicolumn{2}{c|}{Head} & \multicolumn{2}{c}{Pelvis}  \\
    \cline{2-5}
     & PSNR$\uparrow$ & GPR$\uparrow$ & PSNR$\uparrow$ & GPR$\uparrow$ \\
    \hline
    Mono-energetic input   & 33.91  & 24.1588 & 34.95 & 16.4977 \\
    \midrule
    UNet3D              & 57.84 & 99.0762 & 50.51 & 74.8951 \\
    ResidualUNet3D      & 57.94 & 99.3584 & 51.78 & 84.8667 \\
    UNETR               & 58.11 & 99.3597 & 50.02 & 75.9518 \\
    SwinUNETR           & 58.29 & 99.3171 & 49.02 & 77.1031 \\
    TransUNetSE3D       & \textbf{58.86} & \textbf{99.6112} & \textbf{52.45} & \textbf{86.7495} \\

    \hline
    \end{tabular}
    \caption{Quantitative evaluation using PSNR (dB) and GPR (\%) of the proposed Energy-Shifting method with respect to different architectures trained on dataset Head and then tested on two datasets Head and Pelvis. The best results are highlighted in bold. The compared methods utilize the same input features, loss function and training/evaluation protocol.}
    \label{tab:comparison_architectures}
\end{table}

\begin{table}[ht]
    \centering
    \begin{tabular}{l|c|c|c}
    Architectures       & Parameters & Model size & Exec. time \\
    \midrule
    UNet3D              & 65.86 M & 251.233  & 10.9816  \\
    ResidualUNet3D      & 113.72 M & 433.821 & 10.8076  \\
    UNETR               & 160.03 M & 610.472 & 12.7666  \\
    SwinUNETR           & 113.57 M & 440.422 & 19.6291 \\
    TransUNetSE3D       & 165.75 M & 632.282 & 13.1340  \\
    
    \end{tabular}   
    \caption{Number of parameters, model size (in Mb) and average execution time in seconds (s) of compared methods.}
    \label{tab:parameter_comparaison}
\end{table}

In order to evaluate the performance of the compared architectures in the context of fast dose calculation, we measured parameter count, model size, and execution time. The average execution time was computed on an NVIDIA RTX A6000 for the Head test set with a voxel spacing of $2\times2\times2 \text{ mm}^3$. The inference of these models strictly depends on the overlap size of the 3D testing patches, which is set to $0.9$, over an input dose map through sliding window inference. We excluded the time required for mono-energetic input generation via GGEMS, as it requires only approximately 2 seconds. For context, conventional Monte Carlo simulation using OpenGATE requires more than 2 hours to generate the TrueBeam reference dose. The results are summarized in Table \ref{tab:parameter_comparaison}. All deep learning methods achieved an inference time of less than 20 seconds, facilitating near-instantaneous dose calculation. While UNet3D remains the most lightweight, other models require more parameters and longer inference times. Notably, although the parameter count of SwinUNETR is comparable to ResidualUNet3D, SwinUNETR exhibits the longest execution time due to its higher computational complexity (FLOPs floating-point operations per second). Our proposed TransUNetSE3D model has the highest parameter count, exceeding UNETR by 5 million parameters. However, it remains faster than SwinUNETR in terms of average execution time. This efficiency, combined with its superior accuracy, suggests that the proposed method is promisingly suitable for fast dose calculation.


\subsection{Ablation study of the proposed architecture}
\begin{table}[ht]
    \centering
    \begin{tabular}{l|c|c|c|c}
    \hline
    Architectures & \multicolumn{4}{c}{Testing set}  \\
    \cline{1-5}
    & \multicolumn{2}{c|}{Head} & \multicolumn{2}{c}{Pelvis}  \\
    \cline{2-5}
    Training set : Head & PSNR$\uparrow$ & GPR$\uparrow$ & PSNR$\uparrow$ & GPR$\uparrow$ \\
    \midrule
    ResidualUNet3D (Baseline)     & 57.94 & 99.3584 & 51.78 & 84.8667 \\
    Baseline$+$SE                 & 58.75 & 99.5684 & 52.27 & 85.5630 \\
    Baseline$+$SE$+$Beam Block    & 58.70 & 99.5731 & 51.95 & 86.4524 \\
    Baseline$+$SE$+$Trans. Block  & \textit{\textbf{58.86}} & 99.6080 & 51.75 & 85.2394 \\
    TransUNetSE3D               & \textit{\textbf{58.86}} & \textit{\textbf{99.6112}} & \textit{\textbf{52.45}} & \textit{\textbf{86.7495}} \\
    \midrule
    Training set : Head, Pelvis  & & & & \\
    TransUNetSE3D & \textbf{59.02} & \textbf{99.6377} & \textbf{59.81} & \textbf{99.5629} \\
    \hline
    \end{tabular}
    \caption{Ablation study on the proposed TransUnetSE3D. The best results and the second best results are highlighted in bold and in italic respectively.}
    \label{tab_ablation_models}
\end{table}

In this section, we present the results of an ablation study to evaluate the contribution of each component within the proposed architecture. Using ResidualUNet3D as our baseline, we integrated three core modules: the Squeeze-and-Excitation (SE) block, the Beam block, and the Transformer block. As demonstrated in Table \ref{tab_ablation_models}, the inclusion of SE blocks significantly improves baseline performance in terms of both PSNR and GPR by providing local feature recalibration. Subsequently, we integrated the Beam block, which embeds physical beam parameters into the network's latent space. Although this addition did not lead to a substantial increase in PSNR, it appears to improve GPR compared to the baseline. This improvement stems from the fact that the Beam block explicitly accounts for the specific beam geometry and intensity profile derived from the input parameters. Within the same anatomical dataset, adding transformer blocks to the baseline further increased both PSNR and GPR. However, consistent with other transformer-based architectures, this configuration shows weaker generalization when applied to unseen anatomy. In contrast, the baseline equipped with SE blocks showed superior robustness for the new anatomical testing set. To achieve optimal performance, our proposed TransUnetSE3D synergistically combines the robustness of SE blocks, the patch dependency modeling of transformers, and the prior knowledge of beam parameters into a single unified framework.

Furthermore, Table \ref{tab_ablation_models} illustrates the impact of the composition of the training set on the performance of the model. Although our proposed architecture TransUNetSE3D inherently exhibits superior generalization characteristics compared to baseline models, performance is further optimized through multi-domain training. Specifically, the integration of both Head and Pelvis datasets yielded substantial gains in accuracy across all evaluation metrics. When validated on the Pelvis testing set, the proposed model achieved a PSNR of 59.81 dB and a GPS of 99.56\%, a marked improvement over the head-only training set, which yielded 52.45 dB and 86.75\% respectively. This underscores the need of anatomical diversity in training data to ensure high-fidelity dose estimation.

\subsection{Application of Energy-Shifting for prostate radiotherapy}

\begin{figure*}[ht] 
  \tabcolsep=0pt
  \begin{tabular*}{\textwidth}{@{\extracolsep{\fill}}ccc}
    \includegraphics[height=5.3cm]{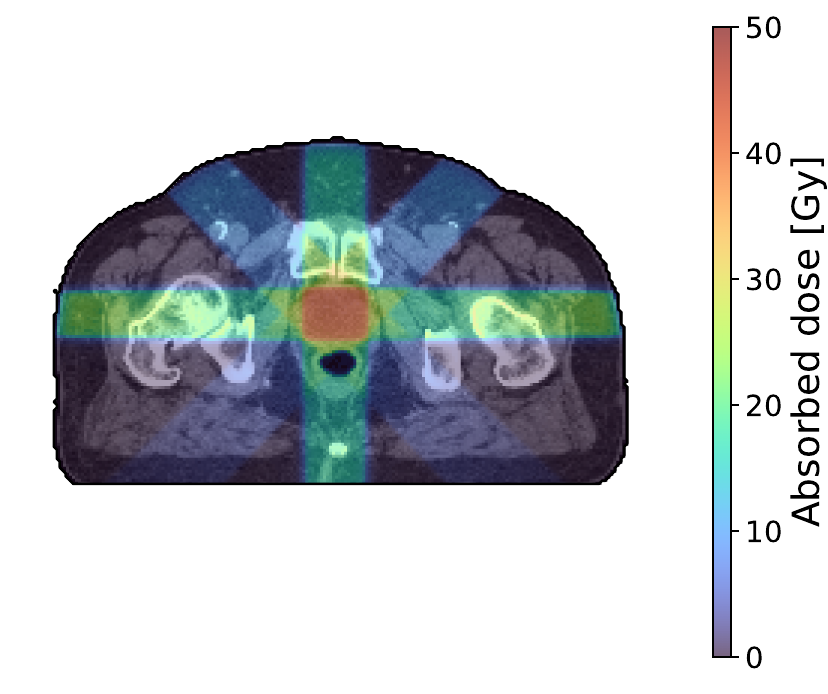}  &
    \includegraphics[height=5.3cm]{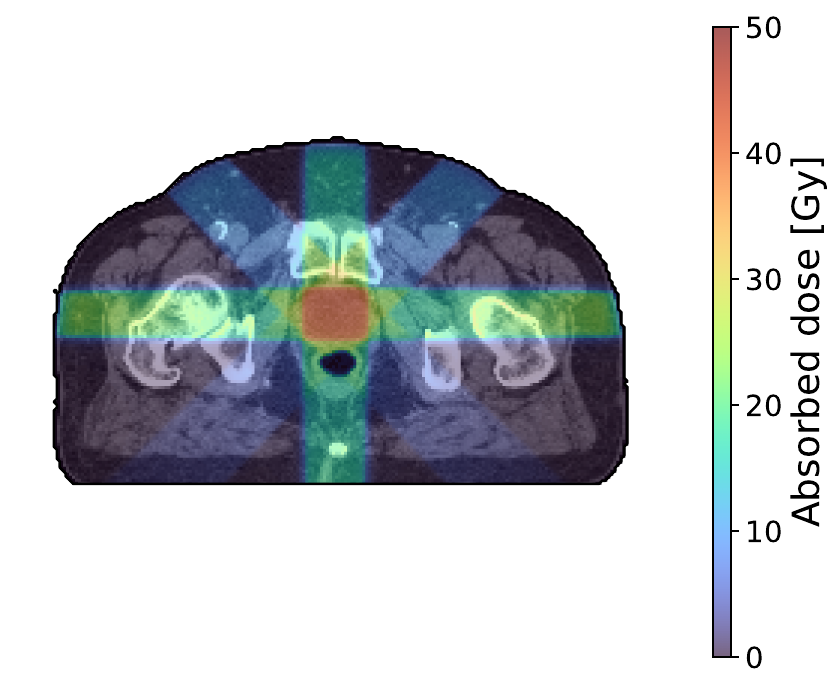} &
    \includegraphics[height=5.3cm]{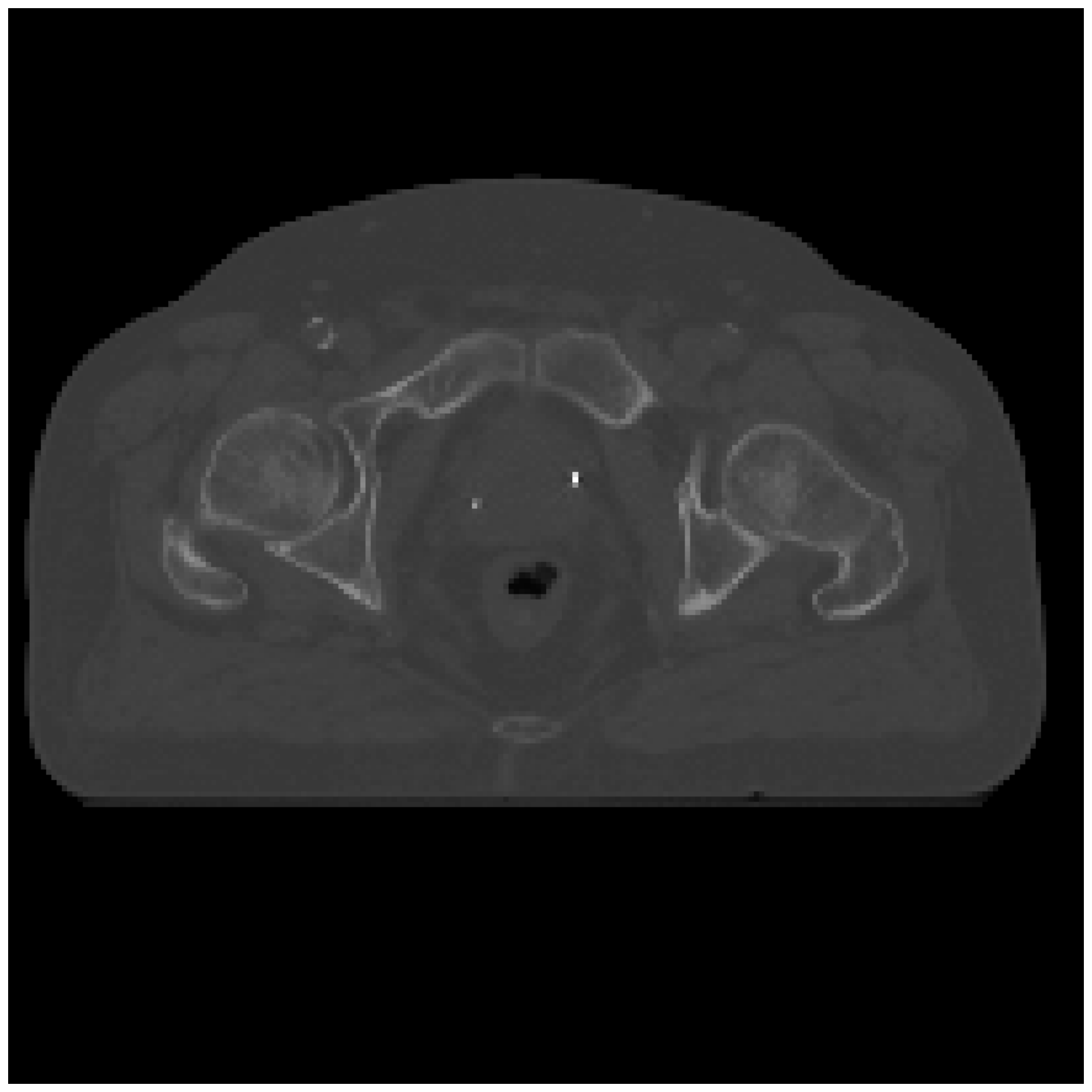}       \\
    (a) MC reference &
    (b) Our dose prediction &
    (c) CT scan
    \\
    \includegraphics[height=5.3cm]{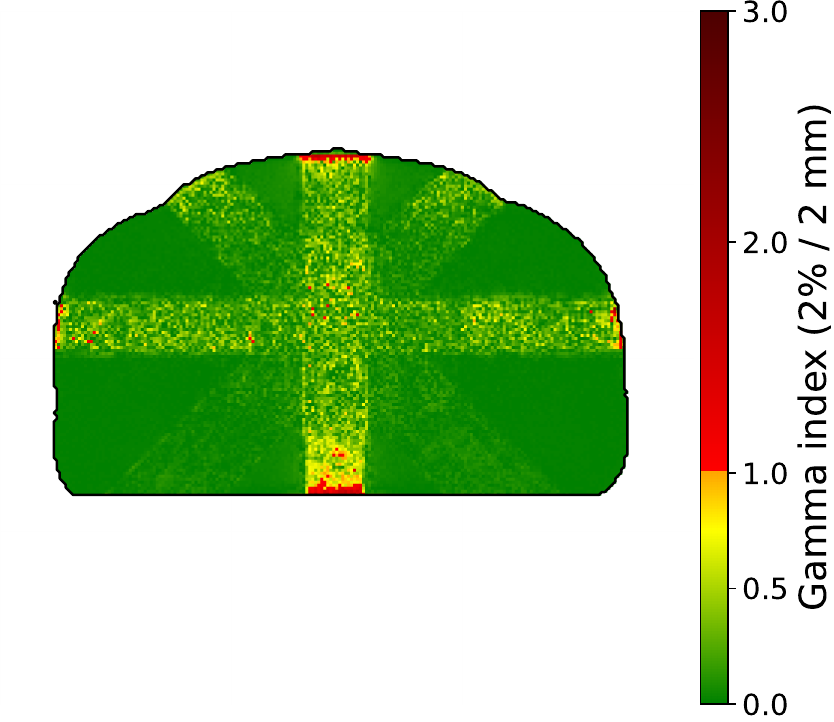} &
    \includegraphics[height=5.3cm]{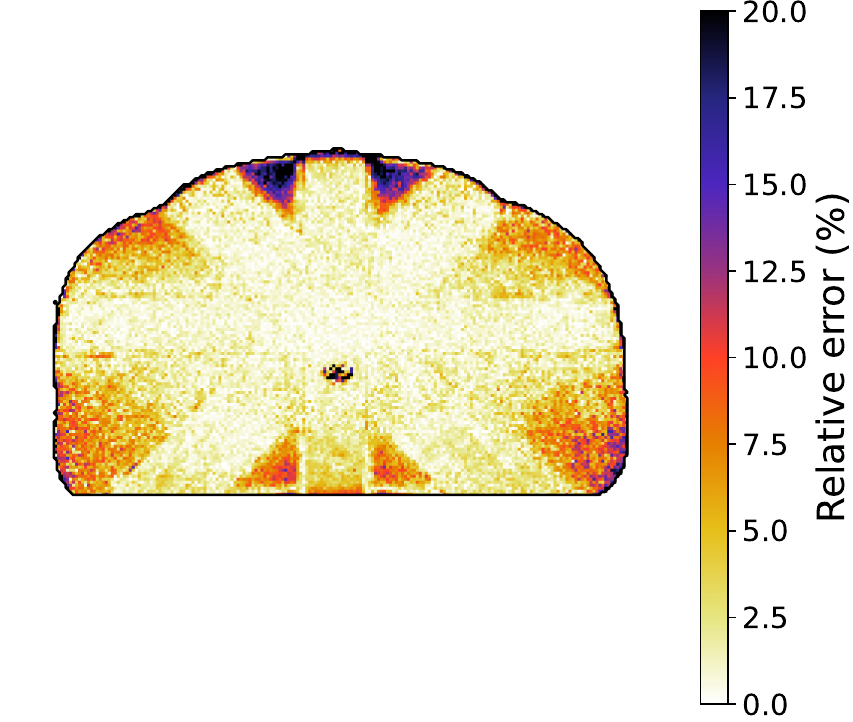} &
    \includegraphics[height=5.3cm]{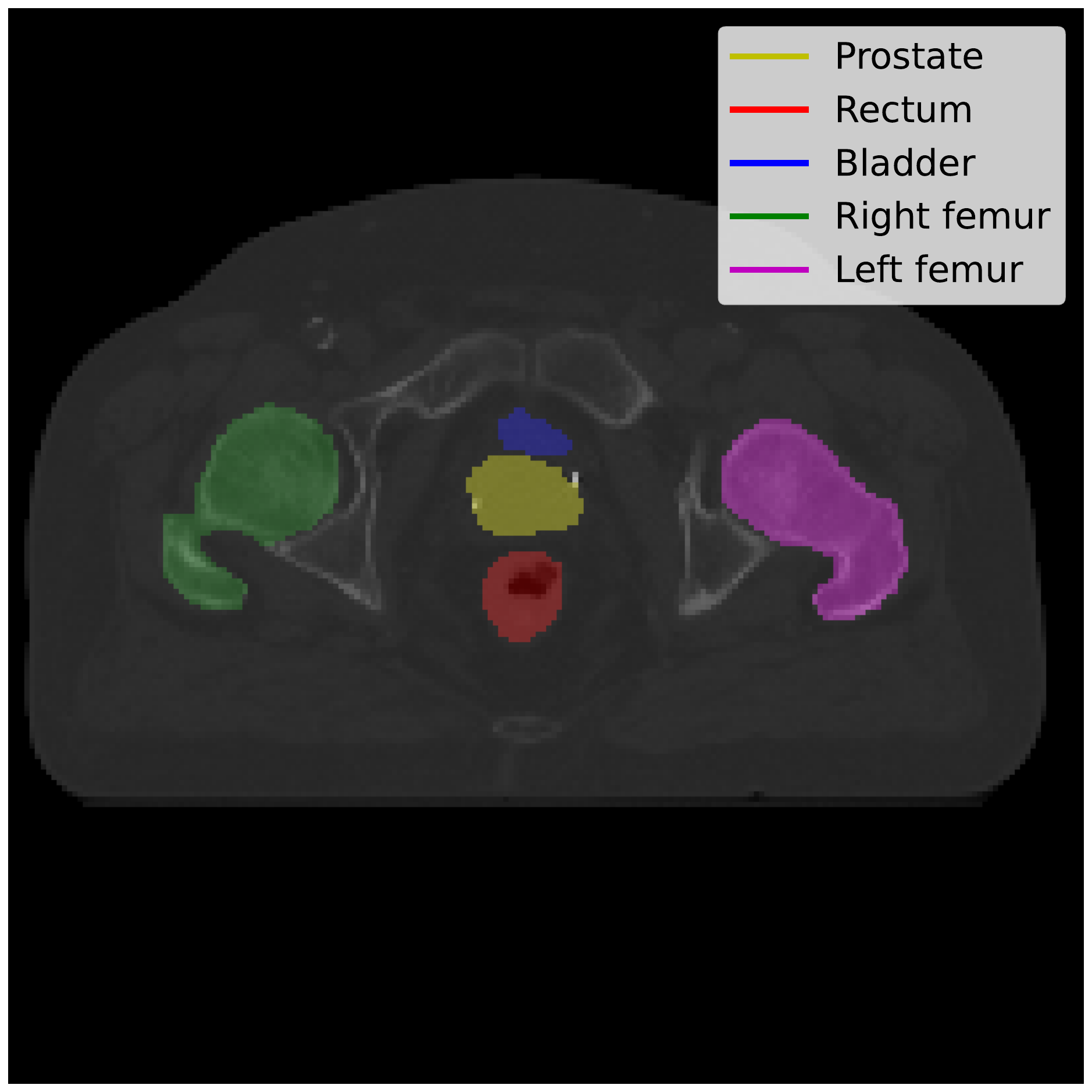}
    \\
    (d) Gamma index analysis &
    (e) Relative absolute error  &
    (f) Segmentation map of CT scan \\
  \end{tabular*}
  \caption{Illustration of gamma index analysis and relative error between (a) Monte Carlo reference and (b) the dose prediction from the Energy-Shifting using TransUNetSE3D. The dose map consists of six beams targeted at the prostate volume, described as (f) the segmentation map of (c) CT scan.}
  \label{results_pelvis_6beams} 
\end{figure*}

In this section, we evaluate the performance of the Energy-Shifting framework within a clinical treatment planning scenario for prostate radiotherapy. The planning geometry consisted of a six-beam configuration with gantry angles at 0°, 45°, 90°, 135°, 180°, and 270°, targeted at the prostate volume.

The MC reference for each beam, utilizing $2^7$ particles, was generated in approximately four hours using OpenGATE on a standard CPU. In contrast, corresponding 500 keV mono-energetic beams were generated in just four seconds using the GPU-accelerated GGEMS framework. Our proposed model subsequently performed the Energy-Shifting inference , using TransUNetSE3D trained on two sets, in approximately 115 seconds. The increased latency compared to previous tests is attributed to the larger matrix size of the pelvic region (relative to Head datasets), requiring the use of more sliding windows to process the entire 3D volume. 

The six independent dose maps were aggregated to generate the full 3D dose distribution. All distributions were calibrated and normalized to the mean dose within the target planning volume, effectively mapping the relative intensities to absorbed dose values in Gray (Gy). Figure \ref{results_pelvis_6beams} illustrates the Gamma index analysis and the relative error between our prediction and MC reference with $2^9$ particles. The proposed framework across the entire six-beam configuration achieved an overall Gamma Passing Rate of 98.27\% (3\%/3mm), PSNR of 66.22 dB and the average relative absolute error of 4.86\%  when compared to the MC reference within the patient body. The prostate volume target is characterized by the highest dose concentration, represented in red, signifying successful beam convergence. The vertical and horizontal beam pairs create regions of high dose intensity, visualized in orange, where the two radiation paths in opposite directions overlap. In the anterior region of the patient, the two diagonal beams exhibit a moderate intensity level, depicted in yellow. Finally, the green-shaded regions represent low-level dose distributions, primarily consisting of scattered radiation and the exit dose of these two diagonal beams. Although minor discrepancies were observed at the beam-skin interface, the Gamma map reveals a similarity within the target prostate region. Furthermore, we observe that accuracy improves proportionally with the number of MC particles; this is because increased particle counts reduce the stochastic noise in MC simulations, allowing the model to more effectively align with the underlying physical distribution. 

\begin{figure}[ht]
    \centering
    \includegraphics[width=1\linewidth]{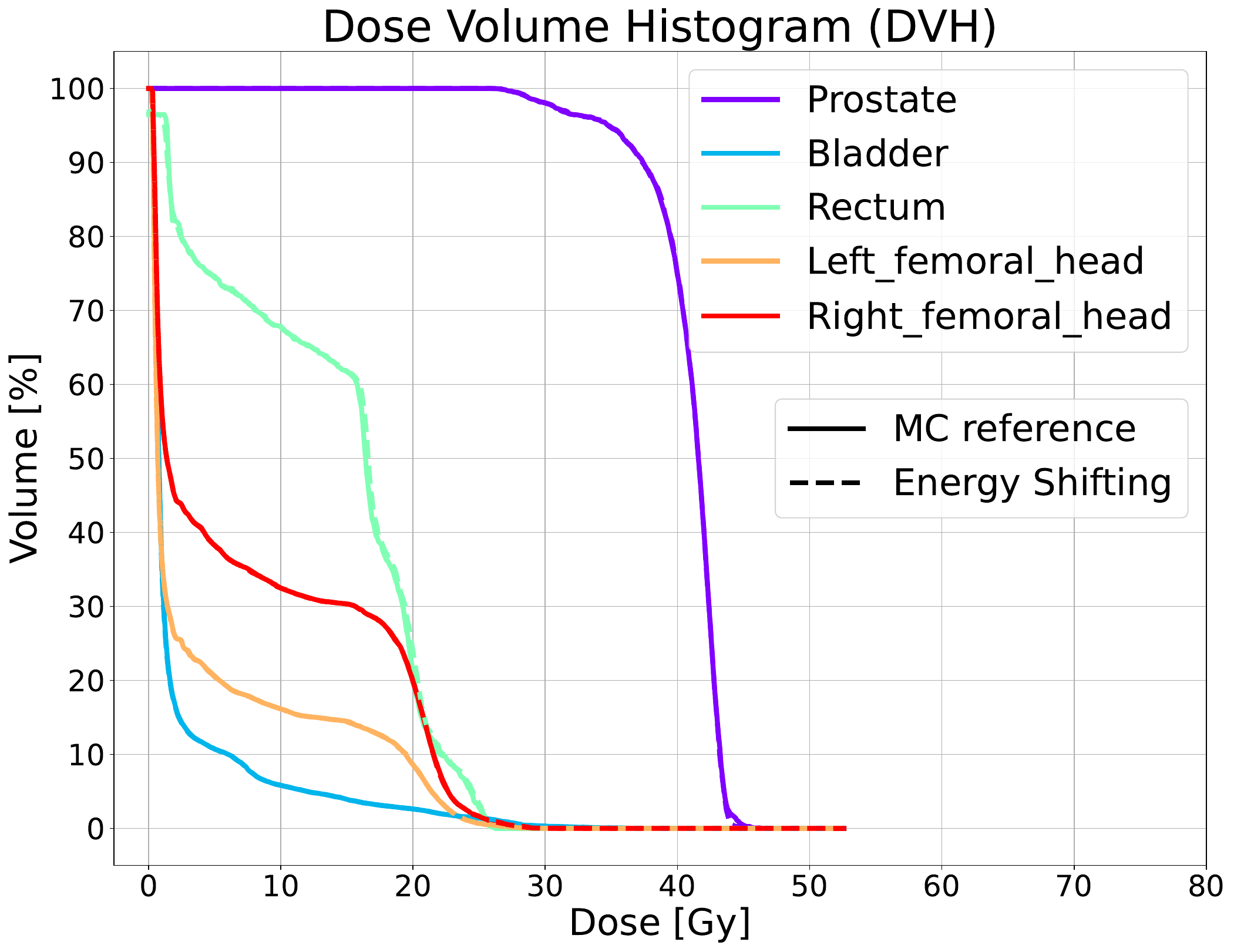}
    \caption{Illustration of dose-volume histogram (DVH) of Monte Carlo simulation and whose of the proposed energy shifting with respect to the organs at risk (OARs) : prostate, bladder, rectum, left femoral head and right femoral head. The DVH is plotted using PlatiPy \cite{chlap2023platipy}.}
    \label{fig:DVH_of_norm_doses}
\end{figure}

Figure \ref{fig:DVH_of_norm_doses} illustrates the Dose-Volume Histogram (DVH) comparison between the proposed Energy-Shifting method and the MC reference for the primary target and critical Organs At Risk (OARs), including the bladder, rectum, and bilateral femoral heads. The prescribed dose for this treatment plan was set at 42.7 Gy. As expected, the Planning Target Volume (PTV) of the prostate receives the maximum dose coverage. Analysis of the OARs reveals that approximately 70\% of the rectal volume receives a significant dose fraction, while the remaining organs show effective sparing with fractional volumes ranging from 10\% to 40\% with half of the prescribed dose. Crucially, the DVH curves for the Energy-Shifting approach and the MC reference exhibit near-perfect dosimetric concordance across the entire volume range for any structure. Furthermore, Table \ref{tab_doses} summarizes the mean dose comparisons, demonstrating that the percentage errors between the two methodologies are negligible. These findings confirm that the proposed hybrid architecture provides a high-fidelity reconstruction, meeting the accuracy requirements for clinical dose verification.

\begin{table}[ht]
\begin{tabular}{l|c|c}
    & \multicolumn{2}{c}{Mean dose (Gy) $\pm$ standard deviation } \\
    \cline{2-3}
   OARs  & Energy-Shifting  & Monte Carlo reference \\
   \hline
   Prostate & 40.8245 $\pm$ 3.1708 & 40.8380 $\pm$ 3.1641 
   \\
    Rectum & 13.7212 $\pm$ 8.0357 &  13.6158 $\pm$ 7.9370 \\ 
   Bladder & 2.3429 $\pm$ 4.8029 & 2.3644 $\pm$ 4.7858 \\ 
   Left femoral head & 4.1969 $\pm$ 7.1156 & 4.2091 $\pm$ 7.1079\\ 
   Right femoral head & 7.6668 $\pm$ 9.0723 & 7.6809 $\pm$ 9.0744\\
\end{tabular}
    \caption{Mean dose calculation (Gy) with respect to different OARs : prostate, rectum, bladder, left femoral head and right femoral head using the clinical planning system for the prostate radiotherapy.}
    \label{tab_doses}
\end{table}

\section{Discussion}

In this work, we presented Energy-Shifting, an efficient framework for accelerated Monte Carlo (MC) dose calculation. The core paradigm involves leveraging deep learning to map arbitrary polyenergetic dose distributions from a corresponding single monoenergetic dose map generated via GPU-accelerated MC simulation. Our model performs fast inference to predict a full-spectrum 6 MV TrueBeam LINAC dose distribution by utilizing the 500 keV map as a geometrically accurate but computationally light input. The integrated pipeline completes the entire full dose calculation in approximately 20 seconds using a single GPU, compared to traditional CPU-based Monte Carlo simulations, which typically require several hours for Head dataset. Despite this significant acceleration, the framework still maintains high dosimetric fidelity, achieving a passing rate of 3D gamma analysis exceeding 98\% (3\%/3mm). While this study validated the approach for photon-based radiotherapy, the underlying principle of spectral mapping is highly extensible and could be adapted for proton or carbon-ion \cite{zhang2022plan,zhang2023deep} in future iterations.

Furthermore, we introduced a hybrid deep learning architecture that integrates Transformer blocks simultaneously with convolutional residual units to capture both long-range global context and fine-grained local connectivity. Physical awareness is maintained through a dual-mechanism approach: the Energy-Shifting framework, which focuses on modeling the spectral transition, and the embedding of beam parameters. By integrating source-specific data, convolutional residual units, and transformer blocks directly into the network's bottleneck, the model is conditioned to respect the deterministic relationship between beam geometry and textures from the image domain. Our model significantly outperforms established baselines while maintaining the fast execution required for real-time clinical workflows. Future work will explore the integration of more advanced attention mechanisms, such as Swin Transformers \cite{liu2021swin}, to further refine long-range spatial dependencies. Beyond dosimetry, the feasibility of the proposed architecture makes it a promising alternative to broader medical imaging challenges, including semantic segmentation \cite{conze2023current} and low-dose post-reconstruction denoising \cite{bousse2024review}.

Our plan to further extend this work is structured into several steps. The first one is to significantly improve the computation time in order to achieve inference times below one second. This initial version will establish a baseline in terms of robustness and high accuracy of the proposed solution. Subsequently, a retrospective clinical study will allow us to evaluate our approach using more realistic treatment plans and to derive large-scale statistical results.
Finally, the last objective of this work will be to develop a dose calculation model for pencil beams or beamlets. The goal is to use this dose engine for inverse planning applications in external beam radiotherapy, thereby enabling consideration of beam collimation using the multileaf collimator (MLC).


\section*{Acknowledgment}
All authors declare that they have no known conflicts of interest in terms of competing financial interests or personal relationships that could have an influence or are relevant to the work reported in this paper. 

\ifCLASSOPTIONcaptionsoff
  \newpage
\fi

\printbibliography

\end{document}